# Developing a system for securely time-stamping and visualizing the changes made to online news content

# Master Thesis

**Presented by**

**Waqar Detho**

**at the**

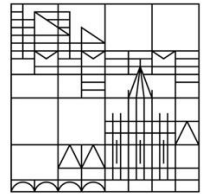

**Department of Computer and Information Science**

1) **Evaluated by Prof. Dr. Bela Gipp**

2) **Evaluated by Prof. Dr Marcel Waldvogel**

**Constance 2016**

## Abstract

Nowadays, the Internet is indispensable when it comes to information dissemination. People rely on the Internet to inform themselves on current news events, as well as to verify facts. We, as a community, are quickly approaching an 'electronic information age' where the majority of information will be distributed electronically and tools to preserve this information will become essential.

While archiving online digital information is a good way to preserve online information for future generations, it has many disadvantages including the easy manipulation of archived information, e.g. by the archiving authority. Online information is also prone to getting hacked or being taken offline. Therefore, it is necessary that archived online news information is securely time-stamped with the date and time when it was first archived in a way that cannot be manipulated.

The process of 'trusted timestamping' is an established approach for claiming that particular digital information existed at a particular 'point in time' in the past. However, traditional approaches for trusted timestamping depend on the time-stamping authority's fidelity.

Directly embedding the hash of a digital file into the blockchain of a cryptocurrency is a more recent method that allows for secure time-stamping, since digital information is stored as part of the transaction information in, e.g. Bitcoin's, blockchain, and not stored at a centralized time-stamping authority. However, there is no system yet available, which uses this approach for archiving and time-stamping online news articles.

Therefore, the aim of this thesis is to develop a system that (1) enables decentralized trusted time-stamping of web and news articles as a means of making future manipulation of online information identifiable, and (2) allows users to determine the authenticity of articles by checking different versions of the same article online. The system's usability is evaluated by analyzing the difficulties faced by participants in a usability study and improving upon the identified issues.

## List of Figures

**List of Tables**

**Glossary of Acronyms and Abbreviations**

**Satoshi** - A Satoshi is the smallest fraction of a Bitcoin that can currently be sent: 0.00000001 BTC, that is, a hundredth of a millionth BTC [1].

**IPFS** - Inter Planetary File System is a content addressable, peer-to-peer hypermedia distribution protocol. The goal of IPFS is to facilitate a permanent and decentralized method of storing and sharing files [2].

**Summative Evaluation** - it is a usability evaluation method, which is used to find and eliminate usability problems at the end of the system development or design [3].

**Formative Evaluation** - It is a usability evaluation method, which is used to find and eliminate usability problems during the design or development of a system [3].

# 1. Introduction

Information that is available online can change quickly. According to an estimate, in one year, 80% of pages on the Internet are updated with changes and 60% pages are newly added [4]. There may also be different problems, such as unavailable content. After one year on the internet, only one fourth of the links are available.

Web archives play a vital role in preserving digital information. They preserve digital information at set intervals and digitize non-digital information for future generations. As a result, knowledge can be preserved for years to come. However, at the moment, despite web archiving services, the existence of digital information in a certain state at a point in the past cannot be verified or proved by a secure means that is independent of the web archiving service.

## 1.1 Problem Setting

Most information published on the Web claims that it was existent at a certain point in time, for example in the form of an online magazine's provided publishing date. Most news sources nowadays are only digital and not present in a physical form, for instance as a newspaper or magazine. Hence, they could claim particular online information existed at a particular point in time, or they could manipulate it later and claim that it is authentic. Furthermore, content from information sources like Wikipedia is constantly being changed and we have no method to keep track of changes in the content. The content of Wikipedia is of particular interest for the people of different domains.

Currently, there is no way to verify whether a news website's claim to un-manipulated content is legitimate or not, because information available online could be overwritten, updated, hacked, etc. There should be a possibility to preserve online news content and help anyone verify the time at which the online information existed in a certain state. This helps users find out if the information has been tampered with at a later point in time. Since trends are changing towards online news sources, we aim to specifically take 'online news content' into account, such as online journals, magazines, tabloids, legal documents, Wikipedia, etc.

## 1.2 Motivation and Objective

The secure verification of the time at which online news content was generated and later modified is crucial for preserving the authenticity of news records. Offline digital information can be easily time-stamped using different systems online (see page 21 for an overview of existing systems). Using these systems, users can easily time-stamp any digital file that they can access online, or that they own, such as videos, pictures or text files. It is also possible to verify the time-stamps later to prove the presence of the file at the aforementioned time in the past. There is also available downloadable software for such purposes; chapter 2 describes these in detail. However, there is no system available on the Internet that allows its users to specify the online news media they would like to securely time-stamp or preserve and subsequently keep track of their changes. The news production sources are not interested in creating different versions of news articles and often they update news pages without providing this information to their readers.

An example of covert news article modification is shown in the Figure 1-1. Thus, there is a dire need for a system that can keep track of time-stamped news content and monitor it for changes. Therefore, we require a platform that will provide an interface, which allows for tamperproof time-stamping of user-selected online news coverage.

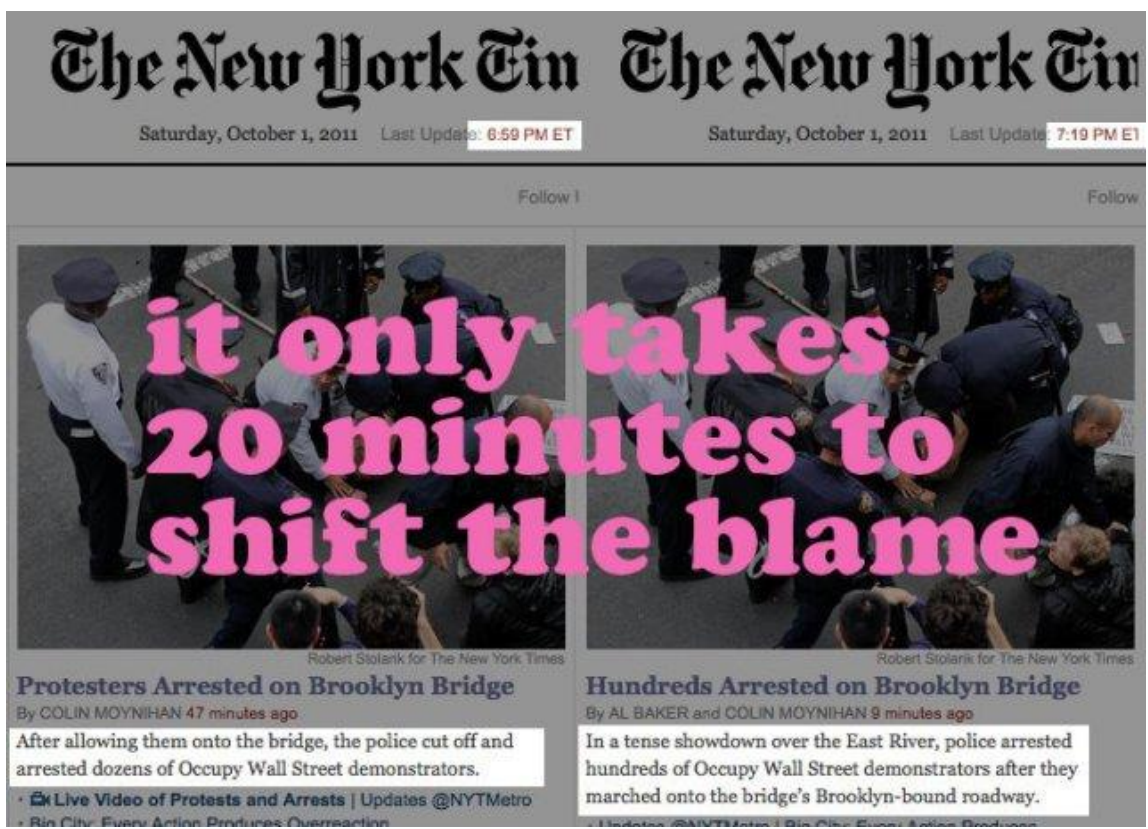

Figure 1-1: An example of a change in a news article within minutes. [5]

There exists sophisticated research in the domain of trusted time-stamping and the details of this research are described in chapter 2. There is a need to find a secure and tamper-proof way to achieve trusted time-stamping services for online digital content. Moreover, it is necessary to implement a trusted method in developing a tamper proof time-stamping system. Thus, the aim of this thesis is to produce a system capable of time-stamping online news articles with the following characteristics:

- A free of charge service for time-stamping online news content.
- Easy to use system for the user.
- A possibility to manage time-stamped content.
- A possibility to get notified if a change in time-stamped content is found.

- Comparing different versions of time-stamping content with each other.
- Automated time-stamping at regular intervals for the online content of interest, as specified by users.

Thus, from the above mentioned characteristics, the main aim of that system is to time-stamp and track changes in the time-stamped content. The research objective is summarized as:

***"Developing a system for securely time-stamping and visualizing the changes made to online news content."***

In order to reach the research objective, we need to perform the following tasks:

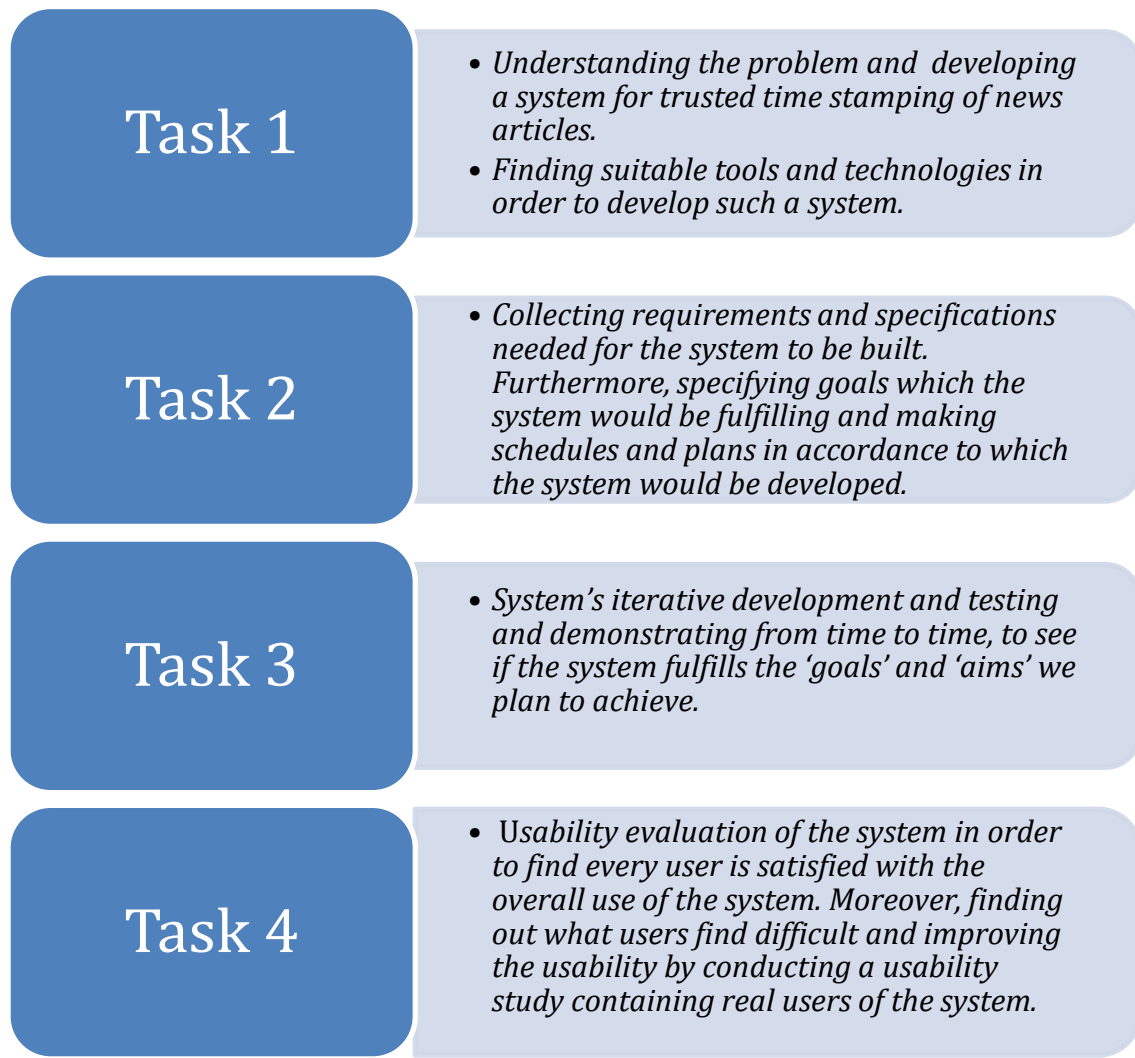

Figure 1-2: Different sub-tasks associated with the research objective.

## 1.3  Thesis Outline

Chapter 1: 'Introduction', discusses the research objective and the need to develop the system researched in this thesis: What are the current techniques available and what is missing or what needs to be improved.

Chapter 2: 'Background', discusses related research in detail. Comparison of available researches with each other. Discussing the advantages and disadvantages of the available research.

Chapter 3: 'Prototype and Use Cases', discusses the developed backend and front end in detail, and the principal use cases and techniques required to operate the system.

Chapter 4: 'Evaluation', presents the evaluation of the developed system with emphasis on the system's usability. The evaluation methods are described and the results are discussed.

Chapter 5: 'Outlook', discusses the future works possible to enhance this system in order to address the system's identified weaknesses and improve the system's functionalities.

Chapter 6 Conclusion

8- The Thesis Appendix includes the usability questionnaires, the results from the usability evaluation, including suggestions for improvement, and a technical documentation of the implementation.

## 2. Background

Web archives play an important role in preserving digital information. Moreover, digital information may also be preserved using trusted time-stamping which securely keeps track of changes in the information that is time-stamped.

### 2.1 Web Archive

Information on the internet can be archived in order to preserve it for the future. There are many existing web archive services for similar purposes, such as archive.org [6] and Pandora Archive [7], which serve as general web archives. Google news archive [8], CNN news archive [9] and BBC news archive [10] serve as news archives. Moreover, online news articles can also be archived on request by available web archives on the internet for example 'Internet Archive' 'Time Travel' [11] and 'Freezepager' [12]. On request, web-archives can archive any kind of digital information, such as files, images, videos, etc. There are different archiving services also available, which can be used as a service. For example, 'University of Michigan' uses archiving services of 'Web Archiving Service', which is now part of 'Archive-it' [13]. Archiving-it is also part of the 'Internet Archive' and has been used by 400 different institutions [14] as an archiving service.

Most online internet archives serve specific purposes and only archive selective information. For example 'Australian Web Archive' for archiving web content which belongs to Australia or '.au' domain [15] and 'Croatian Web Archive' for archiving web content related to Croatia or '.hr' domain. There is a ton of information available on web archives and there are many web archives whose services can be used free of charge. So far, 'Internet Archive' contains more than 10 million text documents and 2.4 million videos and more digital information as shown in Figure 2-1 [16].

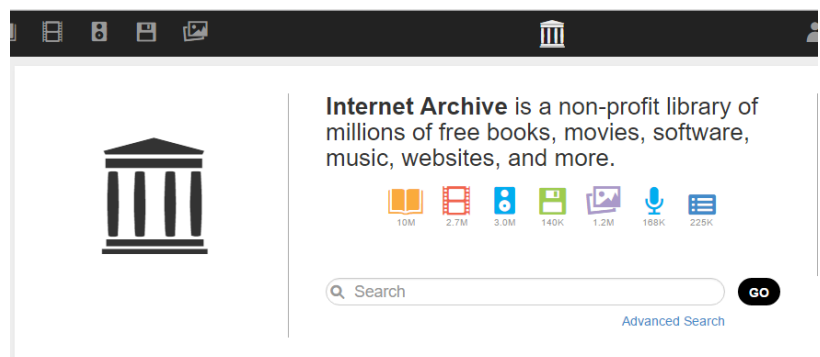

Figure 2-1: Archived information present in 'Internet Archive'

Furthermore, digital information available on the internet can also be archived by providing a URL to the 'on request' web archives systems as shown in Figure 2-2. By providing a URL, one can archive a web-page of personal interest.

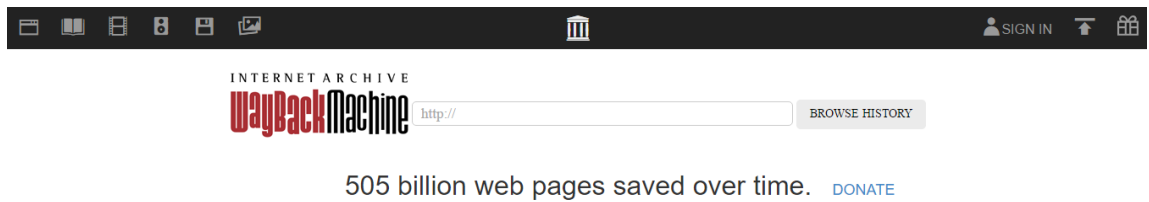

Figure 2-2: Internet archive can be used to archive web-pages by providing the URL of a web page.

Archiving information allows preserving information for the future, and it can protect information heritage on the Web. However, existing internet archives are not secure, because information could be hacked, manipulated by administrators, or face the possibility of server failure. The information preserved is mainly dependent on the faithfulness of the archiving service. Information archives can be proficient for least important data, since we do not want to invest too many resources on it to make it secure. However, data, which is of significant importance for someone on the internet, needs to be preserved in a secure, sophisticated and reliable way.

### 2.1.1  Archive system for news articles comparison

'NewsDiff' is a news archive service, which is used to compare different online versions of news articles. Initially, it supports 'nytimes.com', 'cnn.com', 'politico.com', 'washingtonpost.com' and 'bbc.com.uk' [17]. 'NewsDiff' regularly crawls through the main pages of these news sources and archives news articles from time to time. Users have options to compare different versions of the archived news articles as shown in Figure 2-3.

Figure 2-3: 'NewsDiff' news comparison option for different news from different news sources. [17]

Table 2-1  shows advantages and disadvantages for archiving digital information; the next section describes trusted time-stamping and its uses for preserving digital information.

Table 2-1: Advantages and disadvantages of existing Web Archives initiatives with respect to preserving digital information

| Advantages | Disadvantages |
|---|---|
| Can store any digital information including online news articles. | Existence of preserved information using web archives cannot be proved in future. |
| Can be used free of charge. | Most web archives (80%) archive only country specific content present on the web. [18] |
| Store different versions of web pages. | Validity of archived content depends on the trustworthiness of archiving service. |
| Can be beneficial for researchers, students in different fields such as journalism, politics and history. | Archived content is not comparable in most cases, except 'Newsdiff' [17]. |
| | Do not create trusted time-stamps of the archived information. |

## 2.2  Securing Digital Information using Trusted Time-stamping

Trusted time-stamping is a process for securely time-stamping digital information and controlling the creation and modification of the information. It is assumed that the time-stamped information is not changeable by anyone in the future. The digital information can be time-stamped by using secure cryptographic methods, which are discussed later in this chapter. The next subsection explains entities involved in the creation of a trusted time-stamp for digital content.

### 2.2.1  Entities involved in digital time-stamping

There are four major entities, which are required in the creation of a time-stamp for any kind of digital information or document.

*a) Requester*

A requester is a person or an organization, which owns particular digital information and requires information to be time-stamped.

*b) Information*

Information or data, which needs to be time-stamped in order to proof its existence in the future.

*c) Time-Stamping Authority (TSA)*

An authority permitted to time-stamp information on request. This authority can also be the verification body for the time-stamped information in the future.

*d) Time-stamp*

The exact time of a time-stamp request noted by the time-stamping authority. This time can also be the time according to the law or the time from the time managing organization.

Figure 2-4 provides an overview of the main flow of the time-stamping process.

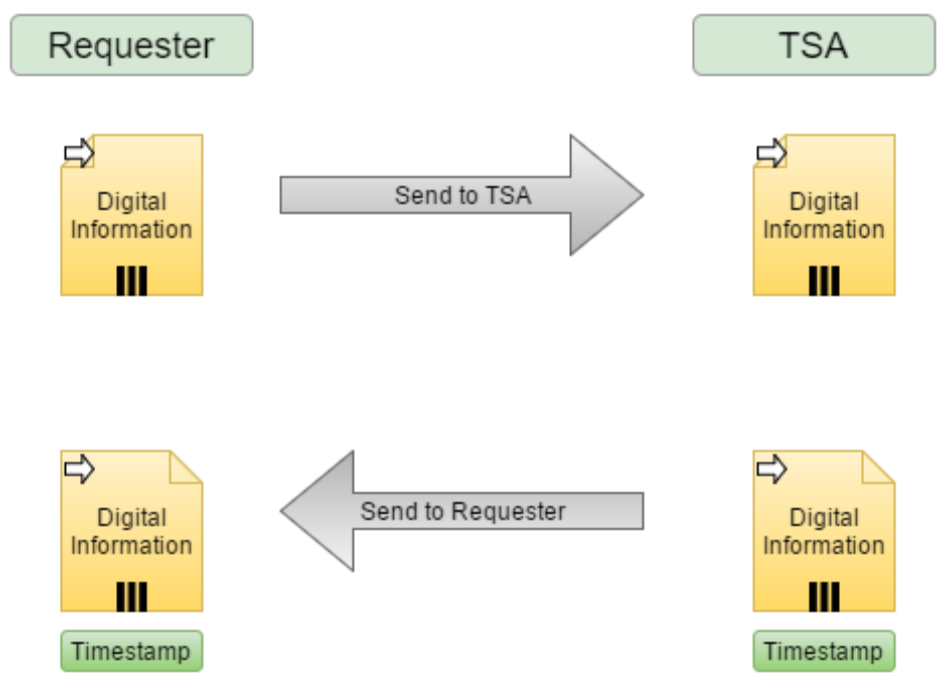

Figure 2-4: Simplest possible way of Time-stamping digital information

The requester sends digital information to the TSA for a time-stamp. The TSA provides time-stamps against the provided information. Now the TSA becomes a witness for the provided information's existence, which can be later proved by using this TSA. The TSA sends the time-stamp, against the given information, to the requester so that the user can prove its existence at some point in the future. This is the simplest possible model for issuing a time-stamp. The next section explains different time-stamping approaches present in current literature.

### 2.2.2   Securing Digital time-stamps

A TSA can emphasize different security goals and there are different ways to provide secure trusted time-stamping services. They can be classified as follows:

*Public Key Infrastructure*

The idea of time-stamping digital documents was first introduced in 1990 [19]. The public key infrastructure (PKI) is the most common security infrastructure used on the internet nowadays for data encryption. The use of PKI in time-stamping systems is outlined by American National Standards ANSI X9.95 [20], Internet X.509 Public Key Infrastructure [21] and International Standards Organization IEC 18014/ ISO standards [22]. There are two important security principals used in RFC 3161, which are PKI -hash and digital signatures.

*Hash*

A hash is a cryptographic function that converts any arbitrary length text into a fixed length text. Hash serves many functions in creating secure systems. In PKI, a fixed length hash is provided to the time-stamping authority in order to obtain a time-stamp so that the time-stamping authority does not get a copy of the text that is being time-stamped. Preserving text at the end of TSA may be a security vulnerability and may result in information leak. This may also result in the information ending up in the hands of someone who is not the owner of the information or a hacker. In addition, Hash is a one-way function and the original text from the hash cannot be computed in case a hash is leaked. It means creating a 'hash' from plain text is possible and creating plain text from a 'hash' is impossible.

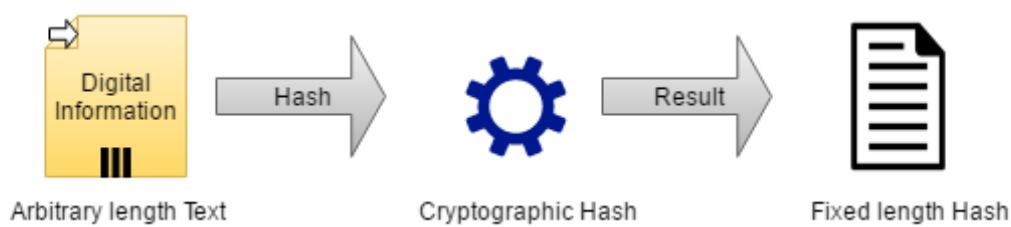

Figure 2-5: If an arbitrary length text is provided to a cryptographic hash function it results in a fixed length text, which is called a cryptographic hash.

*Digital Signatures*

In the online world, digital signatures prove authenticity of the data. Digital signatures use public and private key pairs to encrypt and decrypt data. In nonprofessional language, digital signatures help us find out if the data is from an authentic person. By using digital signatures, data encrypted using a private key of a person can only be decrypted using the same person's public key. In trusted time-stamping services, time-stamped information may be verified only if it is time-stamped by an authentic TSA. For example, in time-stamping process, the trusted time-stamping authority digitally signs information using their private key. The verifier can verify the time-stamped information using the public key of the same TSA, which is publically available to anyone. This is explained in Figure 2-6.

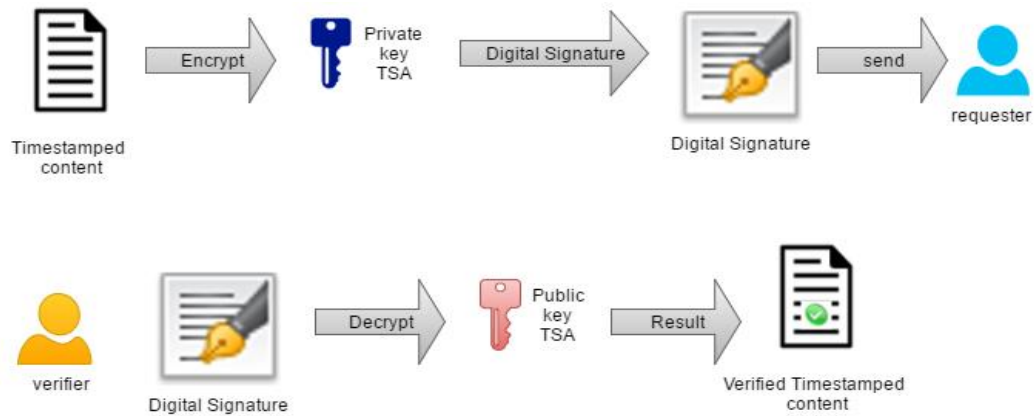

Figure 2-6 : Time-stamping process using digital signatures.

The overall procedure of time-stamping a digital document using PKI is described in Figure 2-7. One cannot time-stamp one's own document, we need a trusted third party in order to prove that our work existed at a certain time point in the past [21]. The use of digital signatures makes it verifiable by anyone using the public key of the time-stamping authority [20]. The hash for the digital content is generated at the requester's end, which is required so that the TSA does not have a copy of the content, which is being time-stamped. The TSA calculates another hash using the first hash (given by the requester) and its time-stamp. Then this hash is encrypted using private key of this TSA and the result of this encryption is called a 'digital signature'. This digital signature is verifiable by anyone using the public key since PKI is a two key / asymmetric cryptosystem.

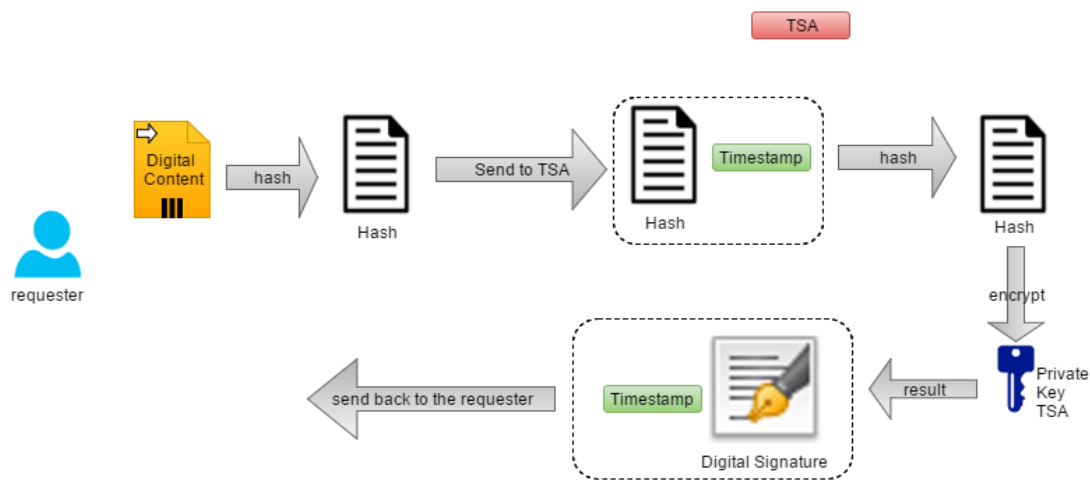

Figure 2-7: Time-stamping process using PKI, which is the most common way of time-stamping digital documents nowadays.

These processes (verification, encryption, decryption public and private keys, etc.) are automatically performed in our web-browsers; it is also possible that someone claims a time-stamped content by using a fake TSA. Here our web-browsers come into play; they only allow trusted sites, which have a valid certificate from a trusted certificate authority. A digital signature is equivalent to a signature with ink and has been used for a long time on the internet to authenticate transactions. The verification process of trusted time-stamping is described in Figure 2-8.

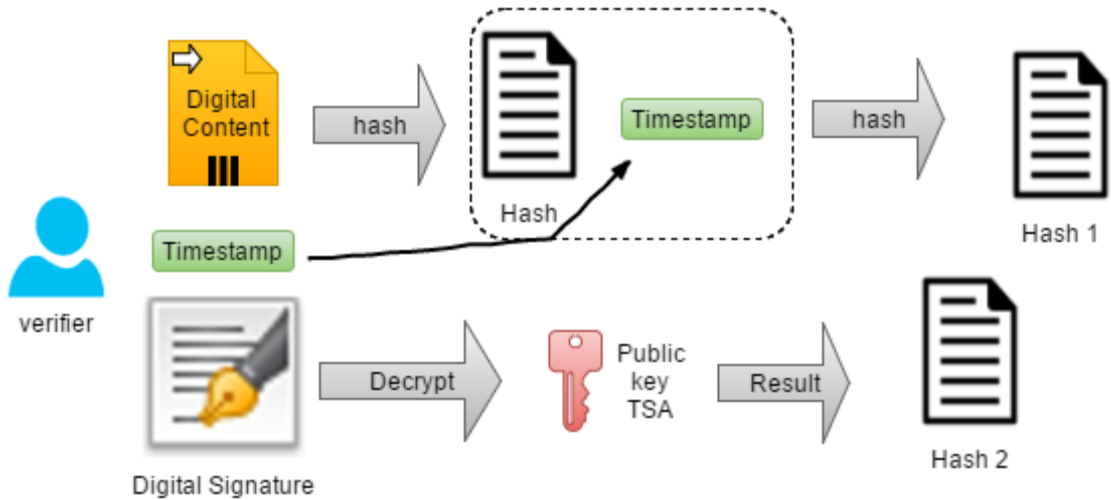

Figure 2-8: Verification process of a trusted time-stamping using PKI.

In order to verify the trusted time-stamp, the verifier needs to calculate the hash of the plain data. As done by the TSA, it also needs to hash the first hash once again with the time-stamp from the past, which is claimed by the user (or owner of the work). Now this hash is called 'hash 1'. The digital signature also needs to be decrypted with the public key, which is publically available to anyone. The result is the 'Hash 2'. If both of these hashes are identical, the time-stamp is verified.

*Using Linked Time-stamps*

Linked time-stamps are based on the idea proposed by Merkle [23]. Linked time-stamps are also described in [19] and [24] with respect to time-stamping digital documents. In regard to approach, all time-stamps are related to each other, making it secure in such a way that altering one hash would require a change in the whole Merkle tree. It is impossible to change an old time-stamp if linked scheme is used. This approach makes linked based TSA an advantage over non-linked base TSA. Linked time-stamps are explained in Figure 2-9.

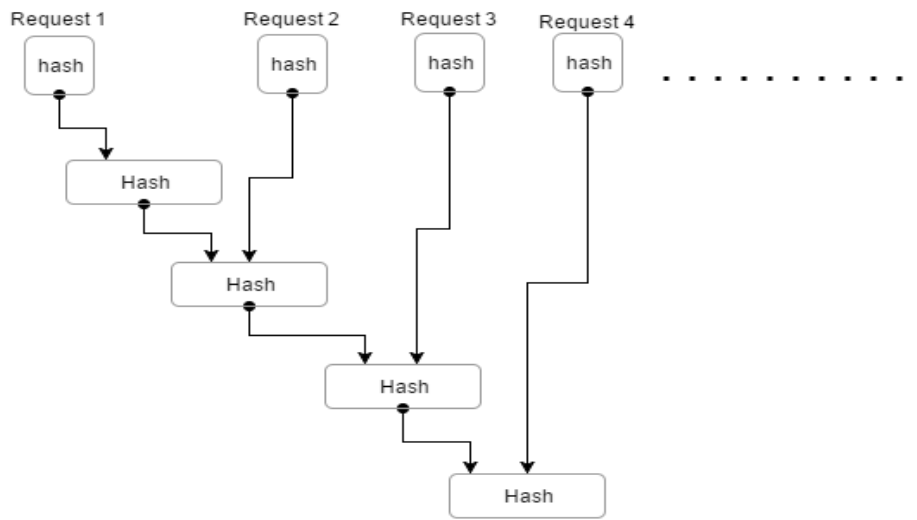

Figure 2-9: Linked based time-stamping scheme.

Any request for time-stamping contains a hash; the TSA needs to add the new hash with the previous hash in order to generate the next hash for time-stamping. This makes the procedure of time-stamping more complex compared to the simple approach discussed earlier. It is also required for the TSA to publish their data on some other trusted systems or make it publically available, for example on an online system or a newspaper. This way, the alterations in the time-stamp data would be impossible, making it secure and increasing requester's faith in the TSA.

*Using distributed TSA*

The issuer of the time-stamp can alter the time-stamp as well as information associated with it [25], as only one single entity is involved in the time-stamping process. In linked based schemes, it is difficult to alter information; however, again it is in control of one single TSA. In order to improve security, more than one time-stamp issuing bodies can be introduced, hence a TSA has no or less chance altering the time-stamping content. This approach is called distributed time-stamping schemes. It is required for distributed TSA that a particular number of TSA share a secret among each other, which is required to generate a time-stamp. If the number of secret data issuers is smaller than a predefined number, a time-stamp cannot be generated. A verifiable digital signature is required, and verification can only be generated if a particular number of distributed TSA add their signatures. Distributed TSA forms a very complex time-stamping system for issuing a time-stamp, since multiple TSA have to work together to generate a time-stamp [26].

*Using Transient Keys*

It is the requirement of American National Standards X9.95 standard [20] to use transient keys during the process of time-stamping [20]. Transient keys are generated during secure sessions and are permanently destroyed after a few minutes. Unlike public-private key pairs which are assigned to a single entity to use it for long time. The data encrypted using a transient key is always associated with a particular time-interval. This can only be used or decrypted for that particular session, since they are dependent on the time at which the data was encrypted using that particular key.

*Using Databases*

A TSA may also save the hashes of the information time-stamped into a secure database, for direct verification of the content by anyone using the internet. In this approach, the verifier only needs to create the hash of the information for which someone claims existence at a particular point in time and match the hash provided by the TSA from their secure database.

*Time-Stamping using MAC*

It is the requirement of X9.95 [20] to use message authentication codes (MAC) in time-stamping processes. A MAC is used for checking the authenticity of messages in information transmission over a secure channel. MAC uses a shared secret key, which is generated during the session generation process. A MAC is generated using shared session keys. The sender transmits the message along with the MAC, which is calculated from the shared secret key. When message arrives at the receiving end, the MAC of this message is again calculated and compared with the received MAC. If both MACs are identical, then there is no change in the message by anyone during the transmission. If the two MACs are not similar, it is possibly due to an attack, e.g. a 'Man in the middle' attack.

*Using Hybrid approaches*

The ANSI X9.95 also explains about using more than one approach for time-stamping process. For example, using 'Linked and Signature' based schemes together. The use of digital signature ensures the data authenticity while time-stamping and verification of time-stamps. On the other end, linked scheme of time-stamps make it difficult to manipulate the content of the time-stamp from the end of TSA. This creates more faithfulness for a TSA. Thus, the more security approaches we use in the development of a TSA the more faithfulness we achieve.

### 2.2.3  Systems for Trusted Time-stamping

*2.2.3.1 Digistamp.com*

This system acts as a trusted time-stamping authority. DigiStamp is not a free service, although they provide a free trial version. DigiStamp follows the RFC 3161 standard [21] for time-stamp creation. DigiStamp is available online and a downloadable desktop application is available as well, for Windows and Mac environments. Besides time-stamping content, they also provide a certificate as an evidence of the content time-stamped [27]. If any modifications have been made with the data, the provided certificate would not be validated, it acts as a digital certificate.

### 2.2.3.2 GlobalSign.com

An online system, which acts as a centralized TSA, provides identity management for digital content. 'GlobalSigns' uses RFC 3161 [21] standards for trusted time-stamping. 'GlobalSigns' uses public key infrastructure PKI for providing identity for everything. Besides providing different security products 'GlobalSigns' also provide time-stamping service for digital documents. 'GlobalSigns' is an online service and time-stamps can be created online using GloabalSign.com [28]. It is not required to be setup, one just needs to make a profile and they offer online access to the TSA. 'GlobalSigns' is not a free service; every individual service needs to be bought against a user created profile in order to start getting time-stamps.

### 2.2.3.3 Freetsa.org

Freetsa is a free service used to create trusted time-stamping services for digital content, against, either a provided hash or a provided URL. When a time-stamping request is submitted, it provides a PDF file against the provided request along with the time-stamp. Freetsa is a centralized TSA and downloading any software or creating an account is not required. Everything can be accessed online and services can be used without providing any identity [29]. Freetsa also provides access to a Restful API in order to perform time-stamping task.

### 2.2.3.4 Truetimestamp.org

Truetimestamp is also a free service for offline digital content. Truetimestamp is actually a trusted TSA. As the requirement of a secure TSA, data is never transmitted, but the hash is transmitted online. The content for the time-stamping could be provided in the form of a hash (SHA 2) as well as digital files. Truetimestamp also provides a user interface for verification of time-stamps either using hash or files [30].

Information preservation using trusted time-stamps is a sophisticated and secure way to preserve digital information. Table 2-2: Advantages and disadvantages of information                                                                      preservation using trusted time-stamping services discusses advantages and disadvantages of information preservation using trusted time-stamping.

Table 2-2: Advantages and disadvantages of information preservation
using trusted time-stamping services.

| Advantages | Disadvantages |
|---|---|
| Digital content can be preserved using secure cryptographic methods of trusted time-stamping. | Most of the services do not preserve or time-stamp online news articles. |
| Information can be verified by using secure cryptographic digital signatures, and anyone in the world can verify it. | The security of a time-stamp depends on the faithfulness of a TSA. |
| May find some free of charge services for trusted time-stamping. | Cannot compare preserved content. |
| Presence of digital information can be proved in a certain time in the past. | Most of the systems do not use linking based schemes or distributed schemes for better trust in information preservation. |

## 2.3 Information Preservation using Distributed Bitcoin blockchain

A time-stamping authority may save the information related to time-stamps i.e. hashes into the database. Moreover, it can also encompass the hashes into the Bitcoin's blockchain network. An example of this work is at [31]. In Figure 2-10 it can be seen that when the hash of digital content is provided to the TSA, it encompasses the hash to a crypto currency along with its time-stamp. Bitcoin is a widely used crypto-currency and, so far, it is found to be secure. Distributed Bitcoin blockchain was introduced in 2008 [32] and it has the highest participating nodes on the internet. The larger number of nodes is considered a security feature as discussed by the creator of this system. It is impossible to generate a Bitcoin attack unless the attacking nodes exceed 50% [33]. Furthermore, Bitcoin uses a decentralized blockchain to store transaction information.

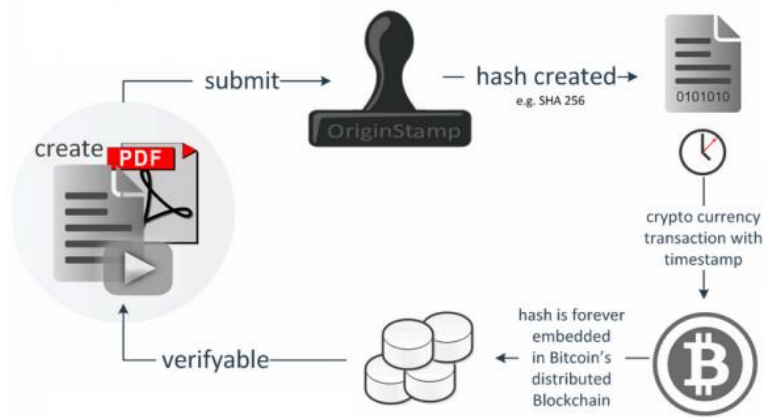

Figure 2-10: Trusted time-stamping using Bitcoin's blockchain (copied from [34]).

This transaction information is tamper proof due to the use of block chains from the Bitcoin network. Furthermore, tampering would require a change in the whole transaction tree (based on the idea proposed by Merkle in 1988 [23]) as the most recent transaction contains the hash of all previous transactions. In order to make this service tamper proof, Bitcoin blockchain has a sophisticated method of storing this information as a transaction by forming a hierarchical architecture by using Merkle tree, and this, in such a way that changing any information would be quite impossible. In addition, Bitcoin stores information of all the parent transactions, providing it with another layer of security and tampering with any information needs a change in all upper levels of the transaction tree. The Figure 2-11explains this procedure in further detail.

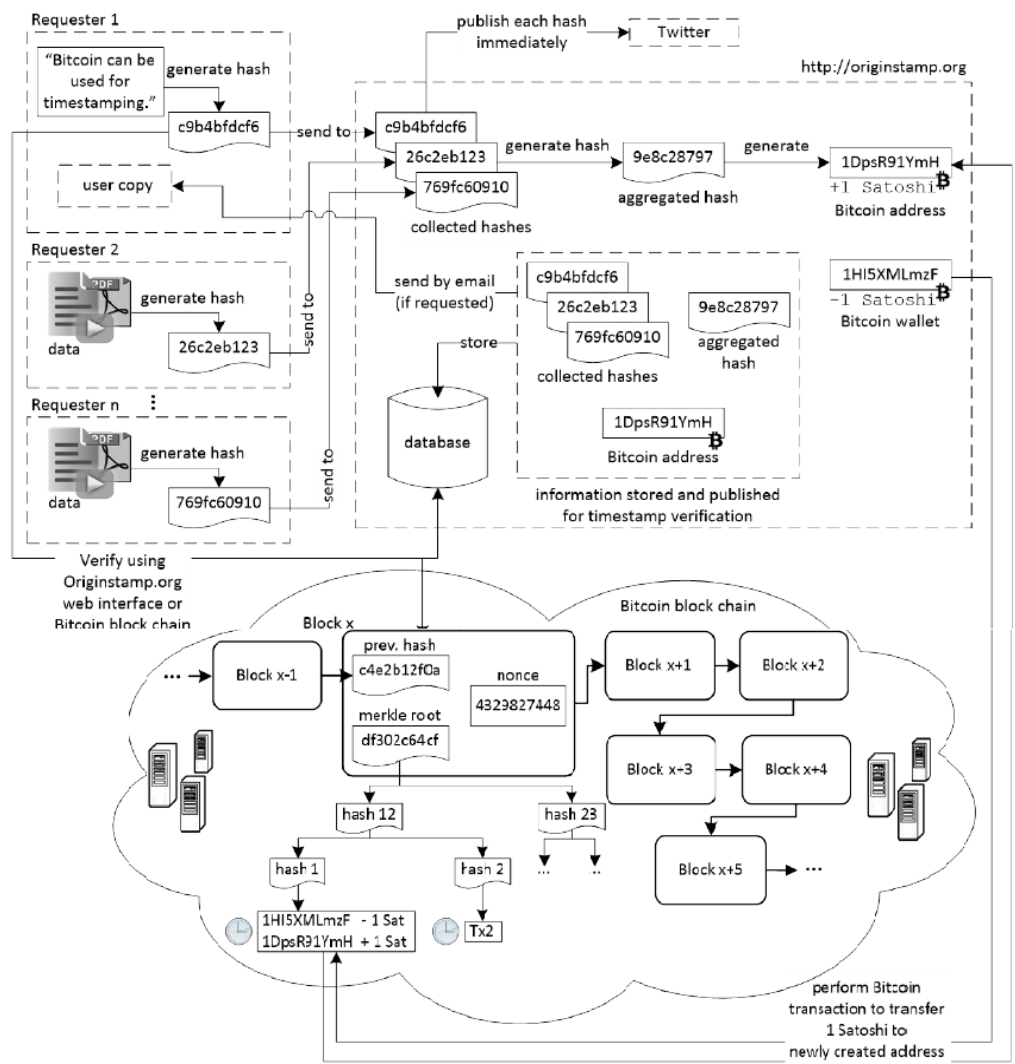

Figure 2-11: Trusted TSA using Bitcoin's blockchain. [31]

When a user accesses 'OriginStamp.org' (Figure 2-11: Trusted TSA using Bitcoin's blockchain. ), the user can upload digital documents for time-stamping. Users send hashes from the digital documents. The hash is created by the client side code, or there are other methods to create a hash of the document and send it to 'OriginStamp'. 'OriginStamp' then creates a second hash from the first hash and its time-stamp. Then 'OriginStamp' creates an aggregated final hash from all the requests of the day and send the final hash to the Bitcoin's transaction network once a day for Bitcoin's transactions and only generates minimal Bitcoin transaction fees, i.e. 1 Satoshi. As discussed before, Bitcoin's transactions are stored forming the Merkle tree, also each Bitcoin's block contains a number called a nonce. This number is formed against the result of a Bitcoin mining. If any change in information in a Bitcoin's block is found, the block does not fulfill a specific security criteria and a nonce is to be calculated again. Calculating a nonce or Bitcoin mining takes a lot of processing and the complete Bitcoin's distributed network is involved in it. Hence using Bitcoin serves another layer of security if a TSA uses it.

### 2.3.1 Systems using Distributed Bitcoin blockchain for trusted Time-stamping

There may be other systems available online that provide trusted time-stamping service. However, the following systems have been found by the result of an internet search.

#### 2.3.1.1 OriginStamp.org

OriginStamp is a trusted time-stamping service for offline digital content for example files, videos and images. As a centralized time-stamping authority may be a security risk, it also uses Bitcoin blockchain network to secure information time-stamped. OriginStamp makes an aggregated hash and encompasses it to the Bitcoin blockchain transaction. OriginStamp aggregates this hash by generating a Bitcoin address with this hash by using Base58 encoding. 'OriginStamp' transfers 0.00000001 BTC for this transaction against. Now this hash is forever embedded in Bitcoin distributed blockchain. This service can be used free of charge, however, information would be encompassed into Bitcoin block chain at the end of the day. If paid for trusted time-stamping, information can be available on Bitcoin soon after the time-stamping request if given. More information can be found on [35]

#### 2.3.1.2 BTProof.com

BTProof is also a trusted time-stamping service for offline digital content for example files, text or directly from hash. BTProof also uses Bitcoin blockchain to secure time-stamped content. BTProof actually converts hash into a Bitcoin's address and performs a Bitcoin transaction using Bitcoin currency. Thus, the information regarding the Bitcoin's address will always remain in Bitcoin's network. If one has the original content from which the hash was created, so one can prove the existence of content at a certain time point in the past through Bitcoin. BTProof is not a free service, $7 is charged for generating each time-stamp. One can also use a personal Bitcoin account to perform time-stamping [36].

Using Bitcoin's blockchain transactions, for preserving digital information is a secure way for this purpose. Bitcoin's blockchain is distributed, information can also be verified, and its existence in the past may be proved in the future, once it is time-stamped. Table 2-3: Advantages and disadvantages of using Distributed Bitcoin blockchain for information preservation. discusses advantages and disadvantages of this approach.

Table 2-3: Advantages and disadvantages of using Distributed Bitcoin blockchain for information preservation.

| Advantages | Disadvantages |
|---|---|
| Digital information can be preserved and hashes of information are only available at the side of a TSA. Hence, it is difficult for a TSA to change any information preserved. | Only for digital content such as files, videos, images, etc. |
| Information saved is distributed throughout the internet, due to the use of Bitcoin blockchain for the distribution network, time-stamps are not vulnerable to manipulation. | Information preserved cannot be compared with other information in the system. |
| There is no limitation of centralized TSA if blockchain is used for example Bitcoin, Ethereum, LiteCoin, etc. . | Do not have different versions of information preserved as in the case of web archives. Since mostly used by people for offline digital content. |
| Very good for researchers or for the people who own some digital work and would like to prove its existence in future. | Online news articles cannot be time-stamped and there is not a single service, which offers this service. |
| Free of charge in the case of 'OriginStamp'. | |

## 3. Prototype and Use Cases

'StampTheWeb' is an online system and can be accessed from anywhere in the world by using a web browser. 'StampTheWeb' is deployed on a web server and contains database and useful files to run the system. Information related to users, time-stamps are saved in a database, and other files (for example PDFs and screenshots) are stored in the same server in the local storage. When a user accesses 'StampTheWeb' and requests a URL to time-stamp, 'StampTheWeb' extracts information of the corresponding URL from the internet and creates the hash for the information as shown in Figure 3-1: shows the deployment view of 'Stamp the Web' and describes how it is operating with other systems..

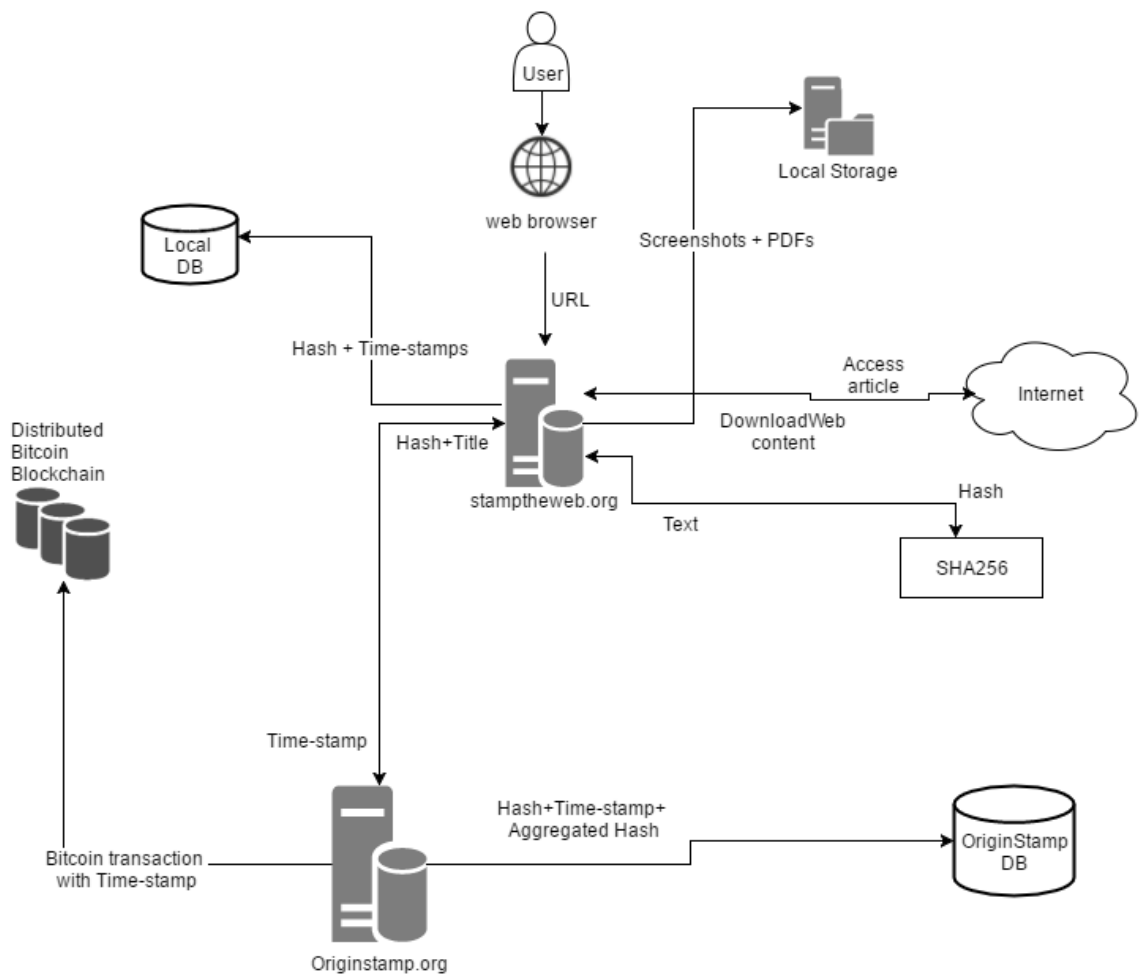

Figure 3-1: shows the deployment view of 'Stamp the Web' and describes how it is operating with other systems.

To get a time-stamp for the hash that represents a news article, the hash is given to 'OriginStamp' API. 'StampTheWeb' also maintains personal time-stamps for individual time-stamps requests in a local database. Each hash along with the aggregated hashes are also maintained in the database of 'OriginStamp' and aggregated to Bitcoin's distributed blockchain by 'OriginStamp' against a Bitcoin transaction [31].

## 3.1   Frontend

The system requires that users create an account, since every news article is associated with a user who time-stamped the news article. The user is able to browse timestamped news articles, but is not able to perform any other action in the system, if the user is not logged in. The available actions shown to the logged in user are shown in Figure 3-2: Overview of 'Stamp the Web' system. The user's personal information is on the top right corner, and title bar presents different options in the system..

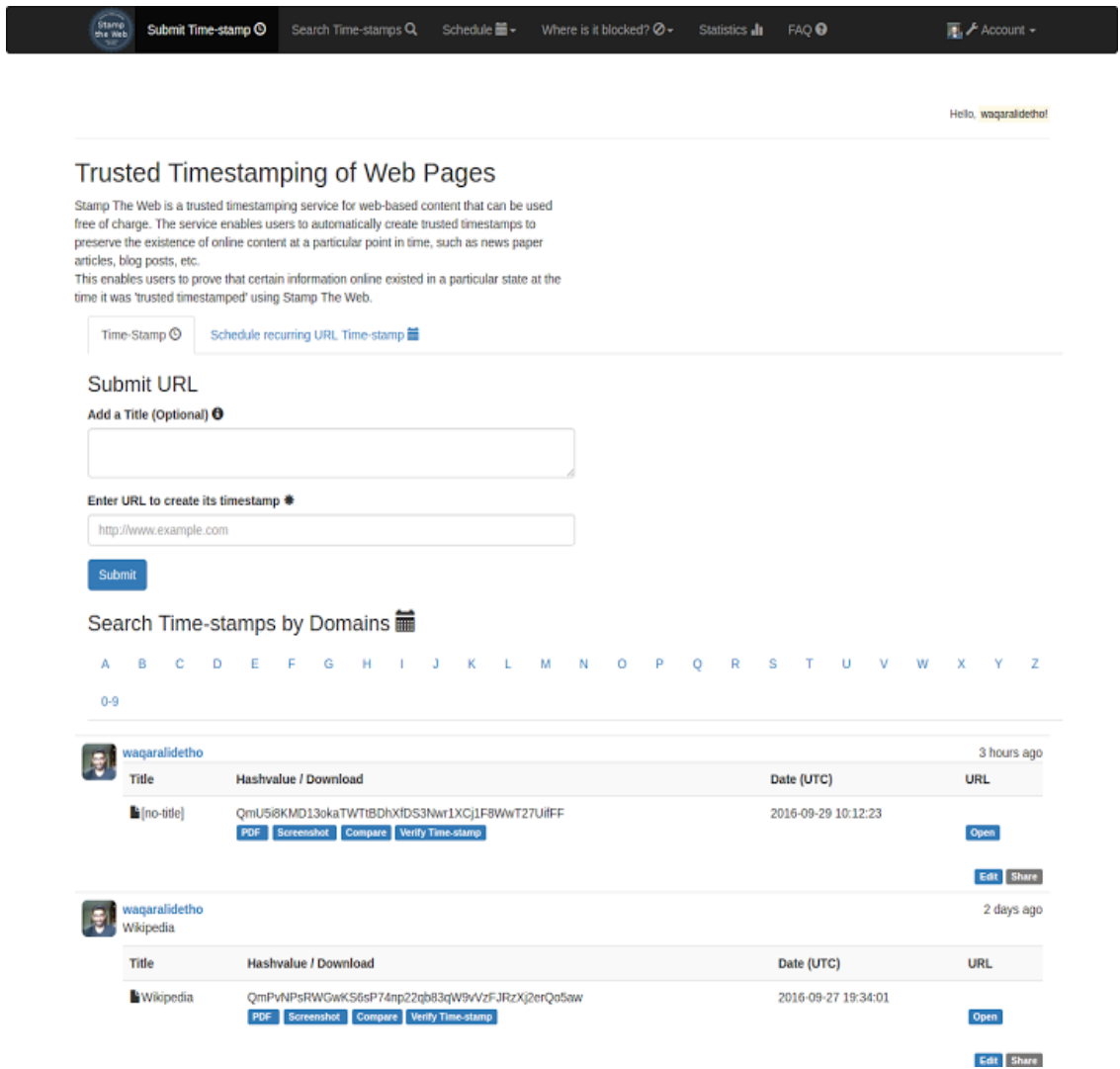

Figure 3-2: Overview of 'Stamp the Web' system. The user's personal information is on the top right corner, and title bar presents different options in the system.

The system considers all time-stamped information as a 'post' in order to distinguish information. As shown in Figure 3-3 a post contains several different things to help improve the readability of a post. A 'post' also contains 'PDF' and 'Screenshot' options to view, so that the user can remember what exactly was time-stamped. A post also contains a hash, which is used to verify the presence of the time-stamp from Bitcoin blockchain.  The 'verify' buttons redirects users to time-stamp and Bitcoin's information from 'OriginStamp'. Furthermore, a 'post contains time-stamp of the post and its title on the web as well as a title assigned by the user.

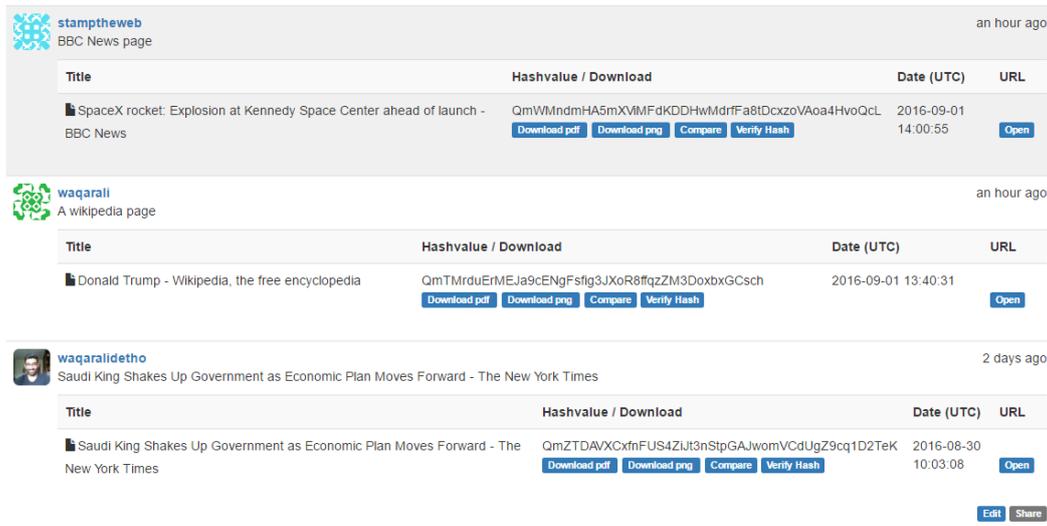

Figure 3-3: Different posts in 'StampTheWeb'

### 3.1.1 Account actions

All users are associated in the system with the selected 'username' while creating the profile. A user can make use of different options related to his or her account and can view personal content.

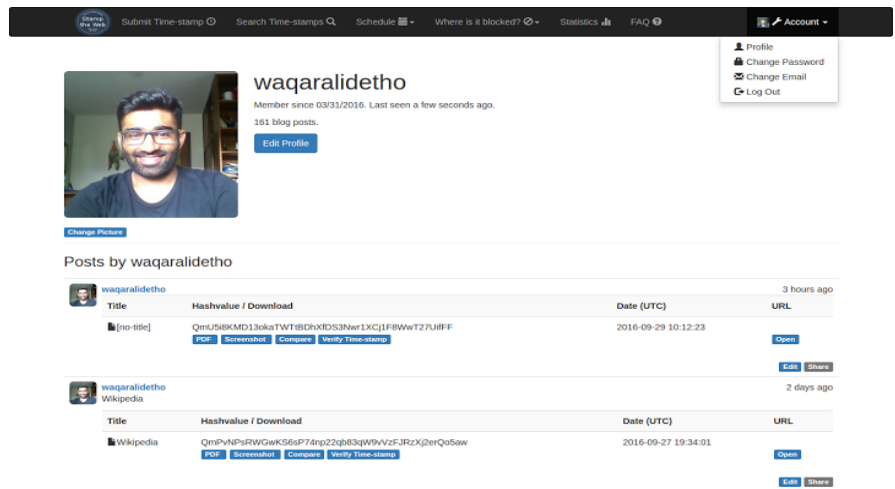

Figure 3-4: 'StampTheWeb' personal profile view and personalized time-stamps.

## 3.2 Core Functionalities

### 3.2.1 Stamping news articles

The system provides a user-interface to time-stamp news articles on the Web. System creates a 'post' against the provided URL, which contains all the information of the post as discussed earlier, along with the time-stamp. It is optional for a user to add a title with each time-stamped post. This will help the user in searching for the article within the system.

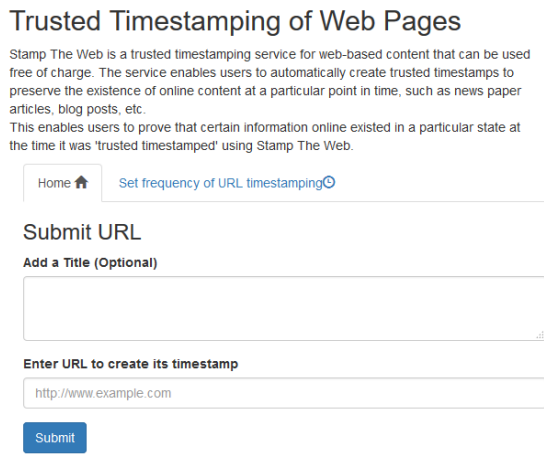

Figure 3-5: Time-stamping a regular news article in 'Stamp the Web' system.

### 3.2.2 Comparing time-stamped articles

'StampTheWeb' offers different ways to compare time-stamps. 'StampTheWeb' may compare the current online version of the time- stamped URL with the same article from another country and another saved time-stamp.

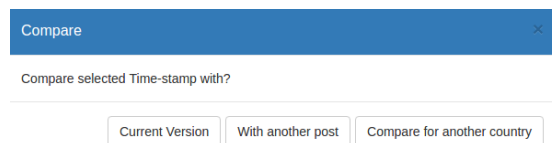

Figure 3-6: Comparing options for a time-stamped article in 'StampTheWeb'.

When it is selected to compare with the same article from another country the system prompts and asks the user to select the second country for the selected post to be compared.

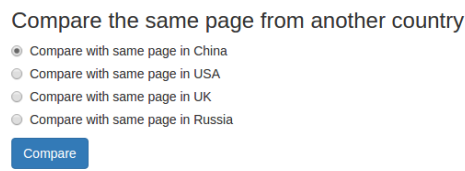

Figure 3-7: comparing an article with the same article from another country.

System compares side by side the selected articles and displays any deletion and additions in a sophisticated manner as shown in the Figure 3-8.

Figure 3-8: 'StampTheWeb' comparing a time-stamped article with the current online version on the web.

### 3.2.3  Searching and browsing saved articles

System provides search feature, which checks three fields in order to find a time-stamp if a text search is performed.

- URL of the time-stamp
- Title of the web article
- Title of the post (if provided by the user)

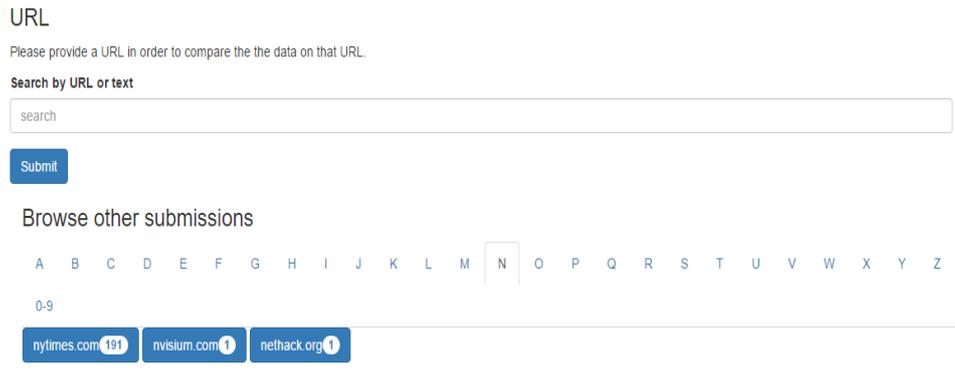

Figure 3-9: Alphabetical browsing of web-articles and searching of web-articles by URL or by text present in the title of the time-stamped post.

Additionally, time-stamped articles may also be searched for by their domain names of the time-stamped URLs. The system automatically creates the list of the domain names of the time-stamped articles. When a search is conducted or a domain name is selected, the system shows the calendar view which contains information of time-stamped posts along with all the time-stamped posts at the bottom. The posts are sorted with respect to date. A time-stamped post may also be opened from the calendar. The calendar view shows different views for example 'Yearly', 'Weekly' and 'Day' view. In addition, 'Month' view as shown in the Figure 3-10.

Figure 3-10: Result of the search query for "world". The system shows all the results in a calendar view as well as each individual post. In the calendar, a green dot shows a time-stamped post, which contains the searched word. The searched word is also highlighted.

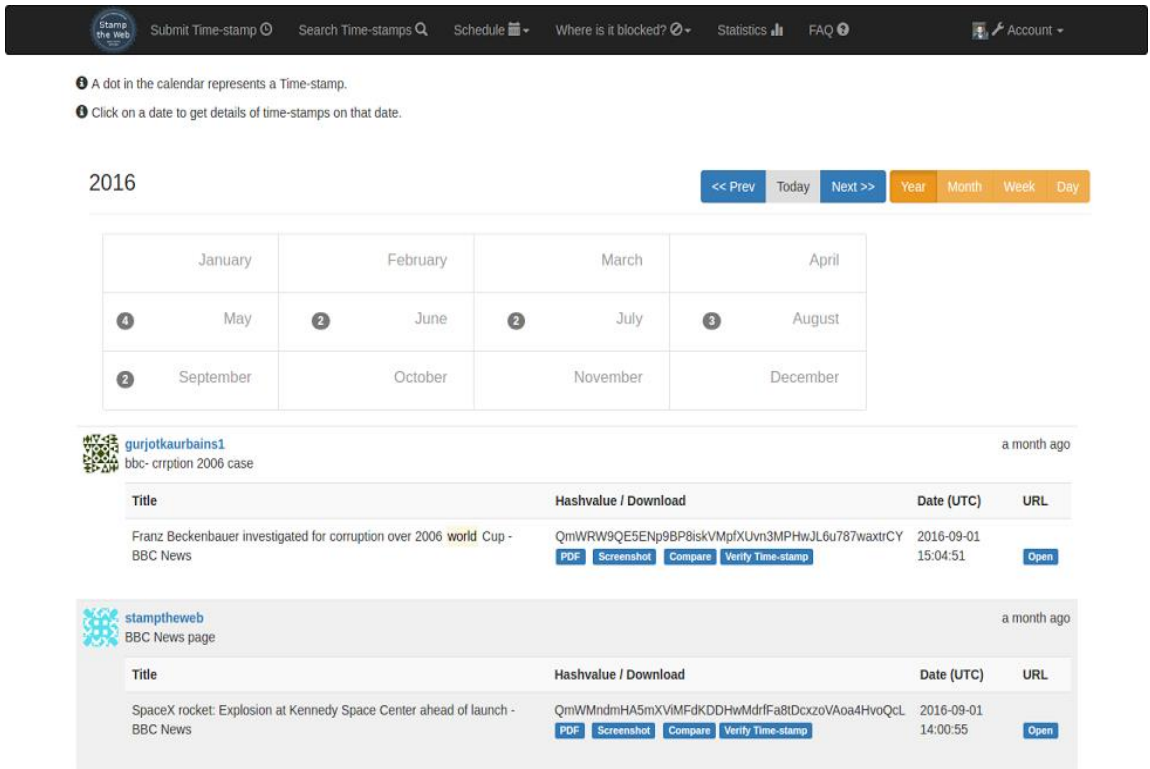

Figure 3-11: Yearly view of the searched or browsed time-stamped articles.

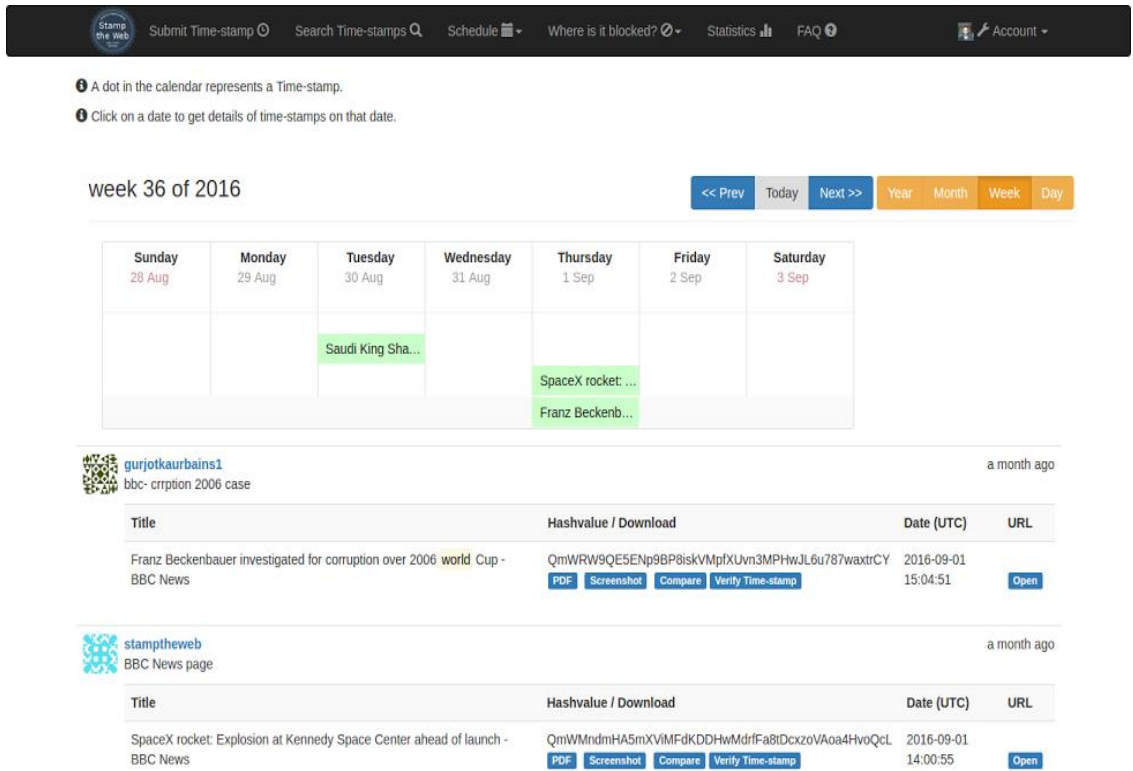

Figure 3-12: Weekly view of the searched time-stamped articles.

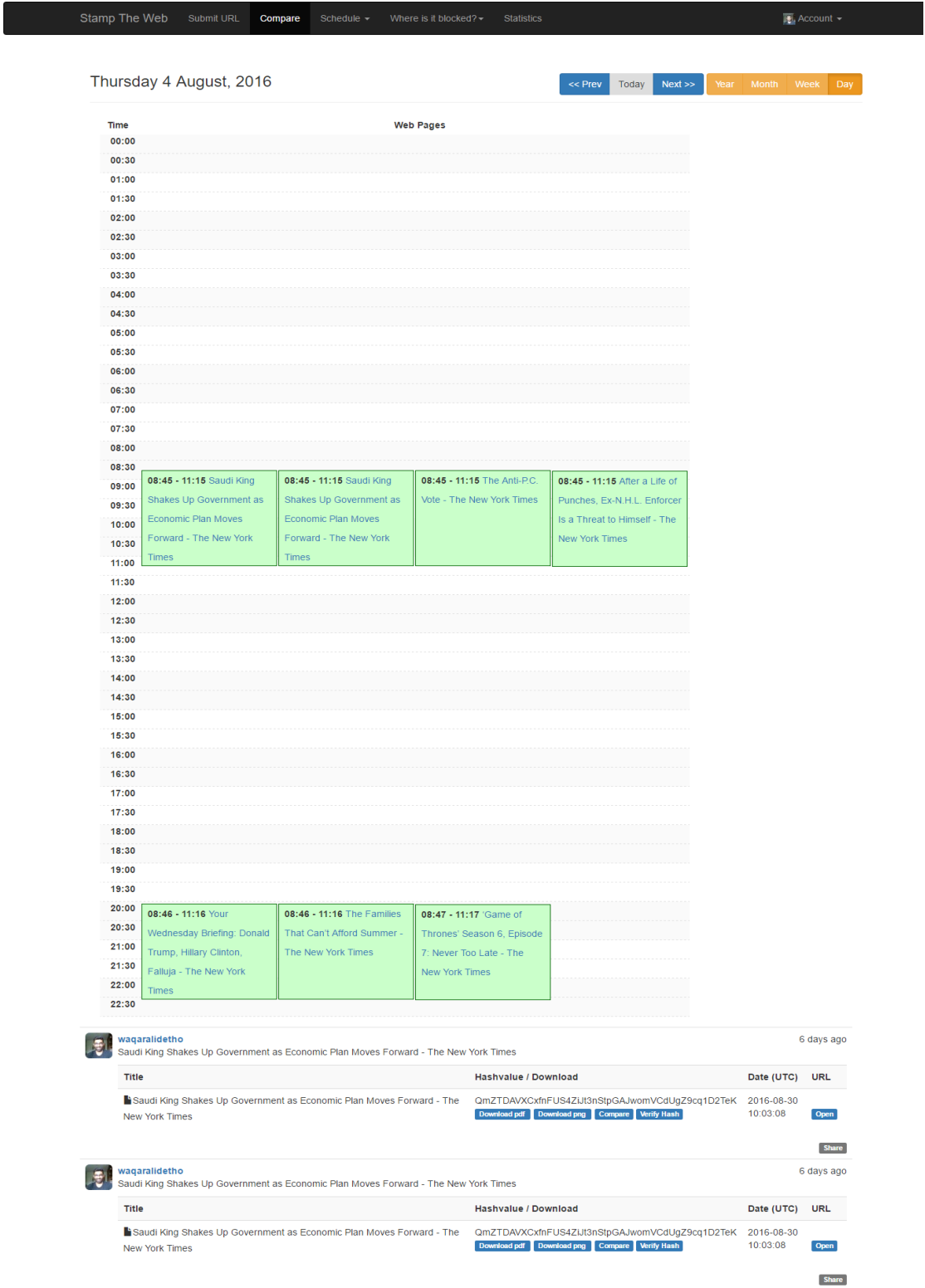

Figure 3-13: Stamp the web 'daily view' of the selected domain i.e. 'New York Times'. The visualization shows the time at which news articles have been time-stamped.

### 3.2.4  Comparing articles from different countries

Time-stamped articles may be compared with the same article's text from another country. Currently from China, USA, UK and Russia. It would be helpful to find out if an article is accessible in a country or not, or if there may be a blockage in certain parts of the text.

### 3.2.5  Check if an article is blocked in a country

The system may also check if an article is blocked in a country. Initially, the support for four countries has been added. The system tries to extract content from different locations by using proxies. If content is accessible, the system shows a message that the provided article is not blocked in the selected country or vice versa. The user needs to provide the article's URL and select the desired country. The result is also saved as a time-stamped 'post' and could be seen by other users. This will help create a platform to find the access of different domains for different locations in the world. 'StampTheWeb' may also help researchers to find about different content blocked in different parts of the world.

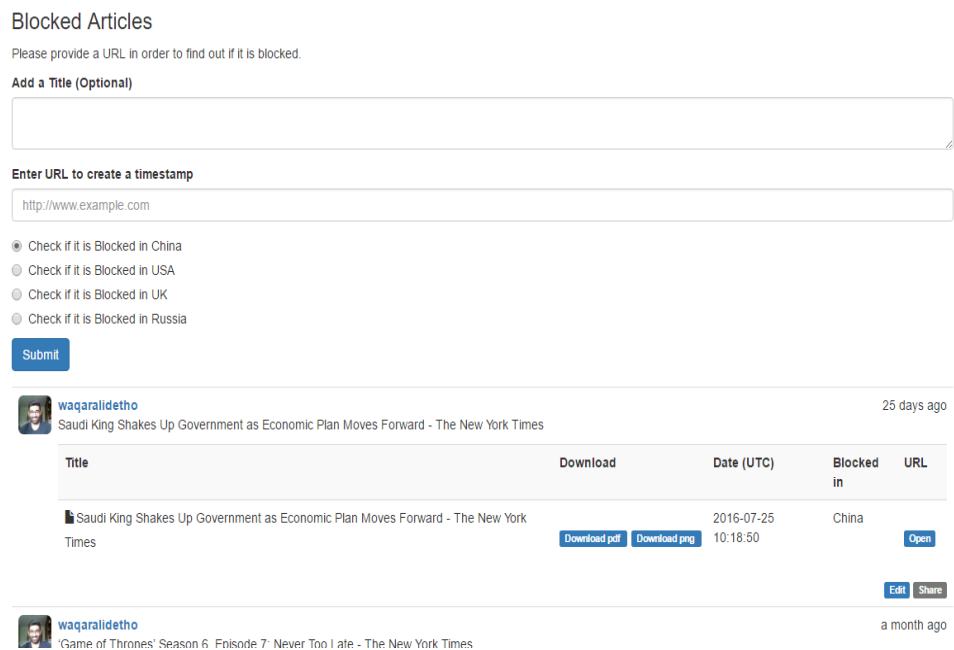

Figure 3-14: Checking where in the world an article is blocked.

If one needs to find out where in the whole world an article is blocked, it can also be checked by going to 'Block Map'. As shown in the Figure 3-15, a Wikipedia page has been checked which is not blocked in any part of the world. Initially, the system checks 31 different countries for generating block map. However, support from the whole world may be added later. In the 'Block Map', the green color means an article is not blocked in that country and red represents blocked articles. This may help researchers and people in the field of journalism to find out about the situation of press freedom in different countries. Moreover, it may help developers of a system discover where their created system is not accessible.

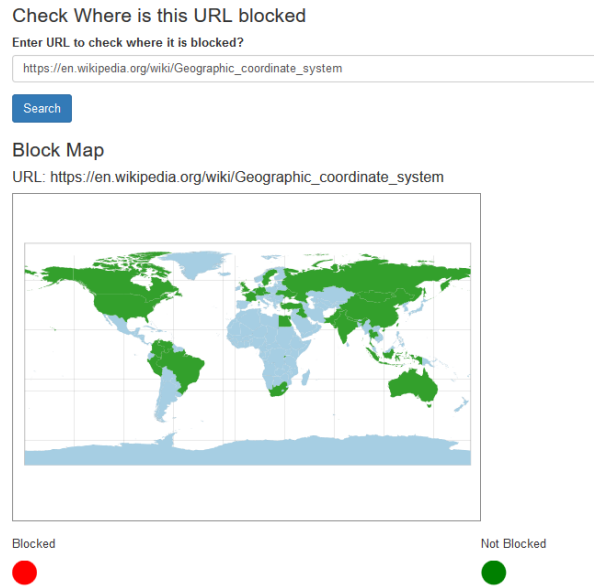

Figure 3-15: Checking where in the world an article is blocked in 'StampTheWeb'.

### 3.2.6 Statistics

The system aims to provide different information to the user regarding system usage in this tab. Initially, the system only shows statistics regarding the most articles used in the system with respect to their country of origin. The more articles are accessed from a country the darker its color would be as shown in the Figure 3-16. The system automatically updates the visualization with use of the system.

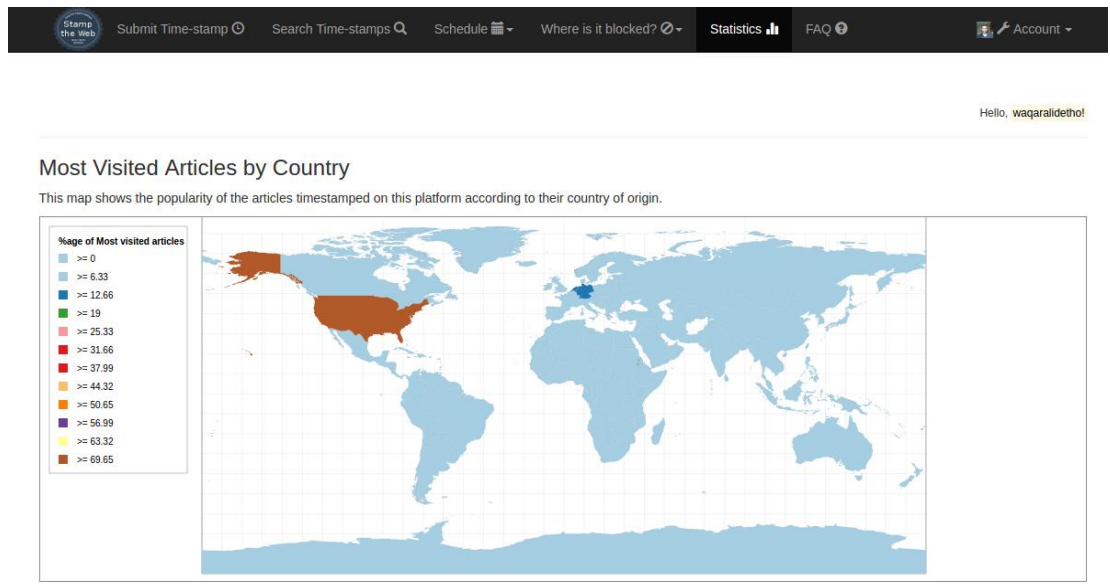

Figure 3-16: usage statistics for the most visited articles from the country of origin.

### 3.2.7  Scheduled Time-stamping

The system may also be used to time-stamp an article with a certain frequency. The system checks the articles frequently according to the given frequency. If the content in an article has been changed, the system would create another time-stamp and notify the user via email (If an email address is provided at the time of time-stamp creation) about the change in the content of the URL.  The system takes 'title', URL frequency and email address, however, only URL and frequency are compulsory fields.

Figure 3-17: Scheduled time-stamping for a provided URL and if any change is found, the system generates another time-stamp for the same article.

'StampTheWeb' may also create scheduled time-stamps for the articles and check the content from another given country. If the content in another country differs from the content of the time-stamped article's content, the system notifies the user by email (if an email was provided). In addition, the system creates another time-stamp for the time-stamped article, if the content is changed. The user just needs to provide a URL, frequency of timestamping, and the country with which the user wants to compare the article.

Submit Time-stamp ⏱  Search Time-stamps 🔍  Schedule 🗓▾  Where is it blocked? ⊘▾  Statistics 📊  FAQ ❓  🔒 🔧 Account ▾

Hello, waqaralidetho!

## Submit URL to be Regularly Checked for Changes in Different countries

**Add a Title (Optional) ❶**

**Enter URL to be regularly timestamped ✳**

http://www.example.com

**The frequency (in days) for which timestamps should be created ✳**

◉ Compare with the Default location
○ Compare with the page in China
○ Compare with the page in USA
○ Compare with the page in UK
○ Compare with the page in Russia

**Notify me in case there is any change in content. (email) ❶**

email@example.com

Submit

Figure 3-18: Scheduled time-stamping for checking if an article is blocked in a selected country. If it is blocked, 'StampTheWeb' notifies the user.

The system maintains the location map automatically. If the user does not provide any location, the system compares the time-stamped content with the default location (in our case Germany, since we have deployed our server in Germany). Figure 3-19 shows the locations used in the system to compare time-stamps.

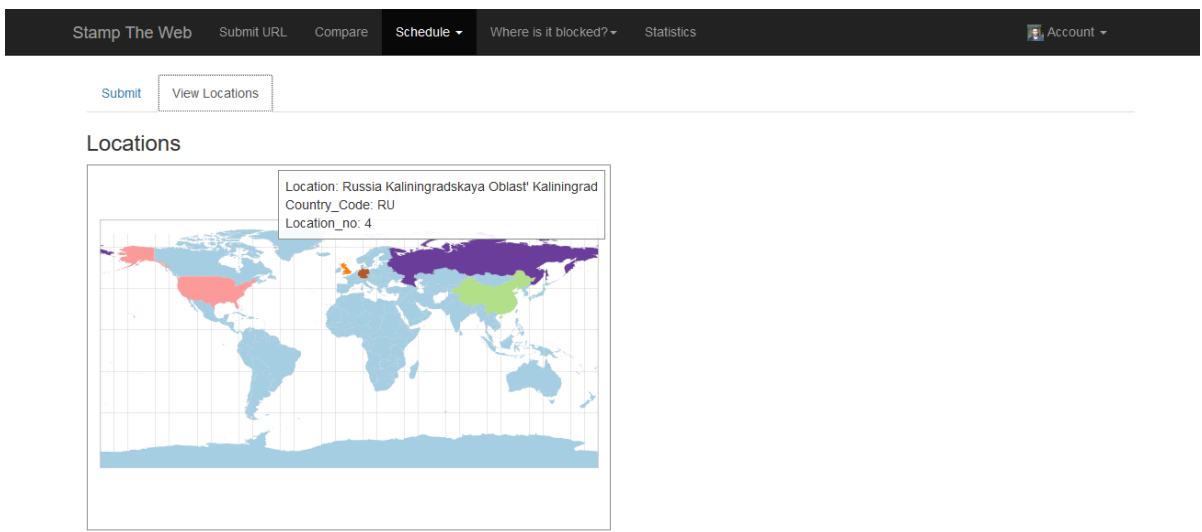

Figure 3-19: Locations used for scheduled time-stamping from different countries can be seen in color in the location map.

If the system finds changes in the content, it notifies the user via the provided email. Figure 3-20 shows an example of an email which the user may receive.

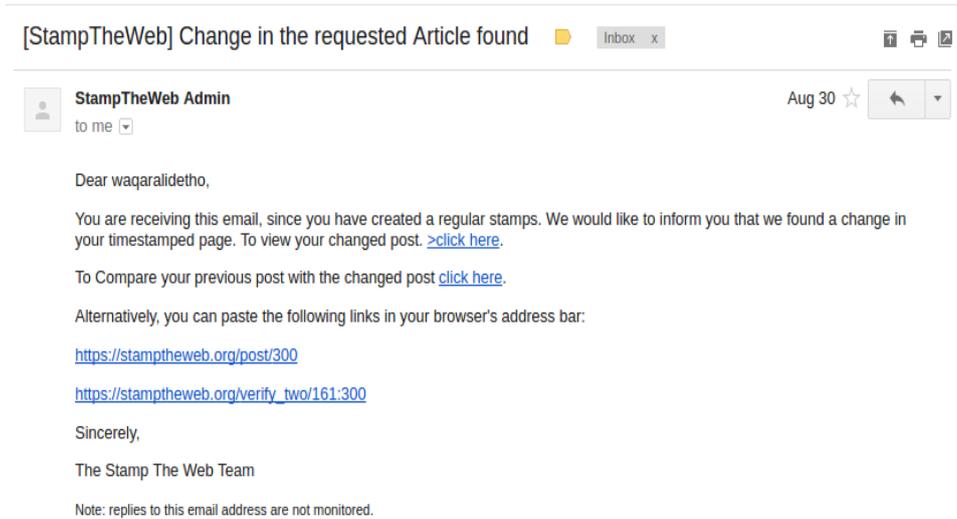

Figure 3-20: Email in case there is a change in content of the time-stamp in scheduled time-stamping.

Similarly, changes in the 'schedule recurring URL time-stamps from different locations' are also notified via email, as shown in Figure 3-21.

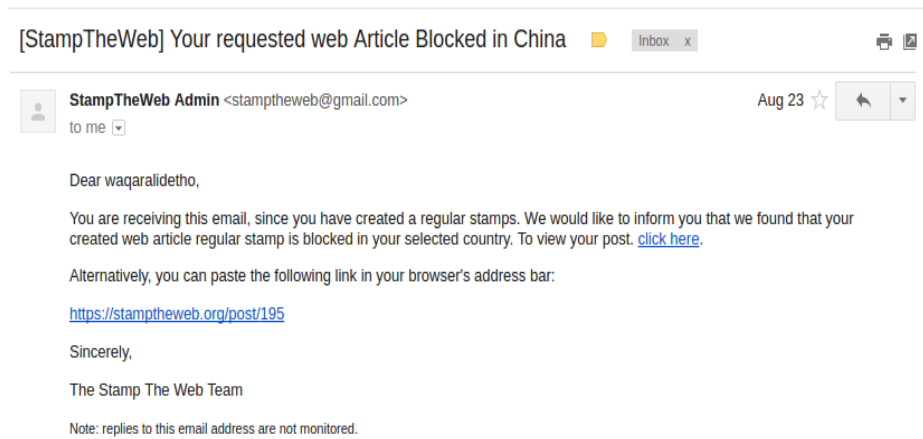

Figure 3-21: Schedule recurring URL time-stamps from different locations.

### 3.2.8  System support

In order to help users, understand the concept of time-stamping and the use of the system, some answers of the basic questions, which may come up in user's mind, have been described in the frequently asked questions, or 'FAQ', page.

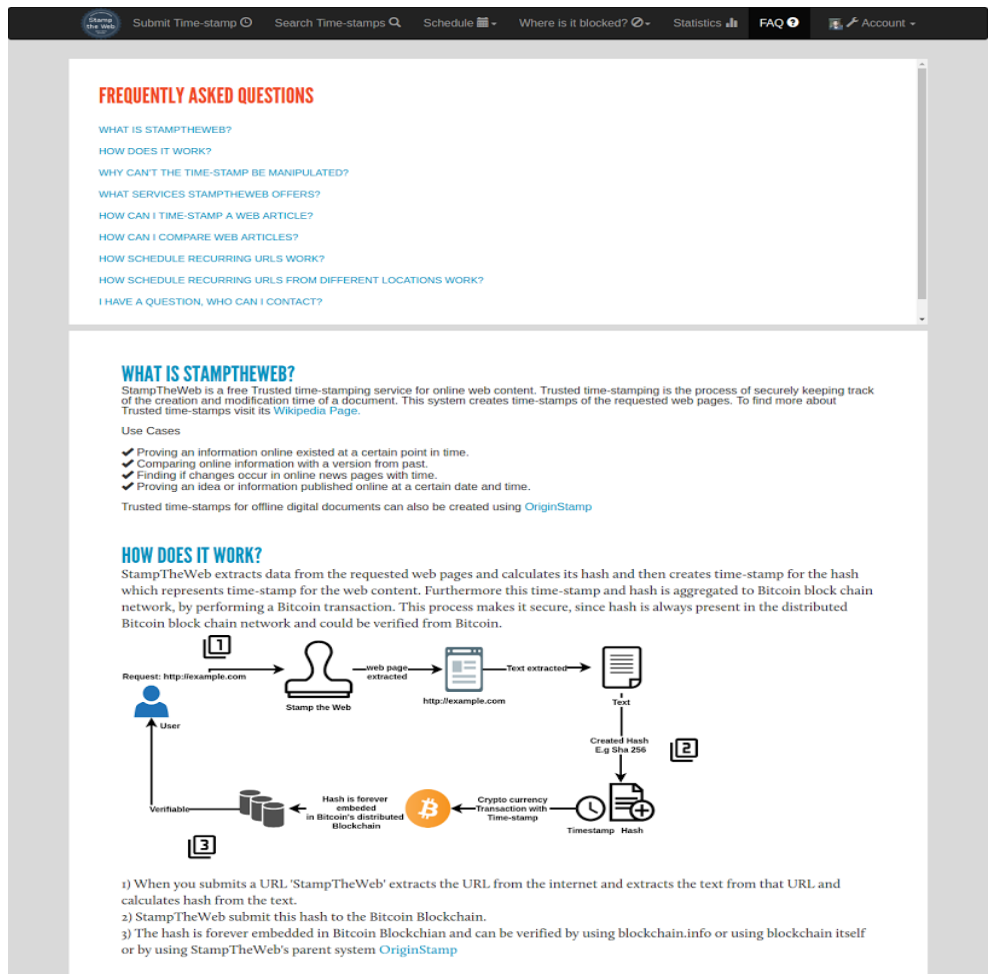

Figure 3-22: Frequently asked questions and answer page

## 3.3 Database Design

Database design and details about each individual table in 'StamptheWeb' are discussed in the developer's documentation section of the Appendix, see page 75.

## 3.4 System Properties

### 3.4.1 Secure User Login

A secure user login is essential to better manage news submissions and created time-stamps that belong to a specific user and prevent information leak or hack. Flask-Login module has been used to make login sessions secure. Using 'strong' protection, the application automatically logs out users if the IP of the user has been changed. Furthermore, token based authentication has been used with a very short expiration (i.e. 3600 seconds) for each token. The client must send an authentication token with each request to the server. As soon as a token is expired, the client must re-authenticate to the system to get a new token. Flask's module 'itsdangerous' has been used to generate and validate the tokens. A user must provide an email in order to register. The user receives an email in order to activate the account, verify the email, and start using the system. A strong password containing at least six characters is required for creating an account.

### 3.4.2 Hash Generation

The system generates and stores a consistent hash of the web articles. The system only considers text and ignores everything else on the web pages like images, ads, etc. The sha256 hash has been used, and this hash is sent to 'OriginStamp' API [35] to get the time-stamp for it. The text on the provided URL is only important for the system.

### 3.4.3 Trusted Time-stamping

The time-stamps are obtained with the help of 'OriginStamp'. 'OriginStamp' is a web service, which provides time-stamping facility to offline digital content. 'OriginStamp' uses the decentralized crypto-currency transactions of the Bitcoin network to store the hashes corresponding to documents and their time-stamps. 'StamptheWeb' generates hash from a web article's text and sends it to 'OriginStamp'. 'OriginStamp' then creates a second hash from the first hash and its time-stamp. Then 'OriginStamp' creates an aggregated final hash from all the requests of the day and sends the final hash to the Bitcoin's blockchain network. Bitcoin is a widely used crypto-currency and it is found to be secure so far. Bitcoin was introduced in 2008 [2] and it has the highest participating nodes on the internet. The larger number of nodes is considered to be a security feature as discussed in [33] by the creator of this system. It is impossible to generate a Bitcoin attack unless the attacking nodes exceed 50%.

### 3.4.4 Display Articles

Display the article in the form of a 'post' along with the title of web article, hash and its URL. A title for the post, besides the web title, can also be added. Title helps users in acquiring articles quickly. Furthermore, the screenshot and PDF are also displayed with the saved news articles.

### 3.4.5 User Profiles

User profiles help distinguish articles according to their respective users. Every saved article in the system is associated with a user. An unregistered user cannot time-stamp a web resource. However, every user can see profiles and posts from any other user. If required, these privacy settings could be changed easily in the future.

### 3.4.6 Scheduled Time-stamping

A user can schedule the time-stamping of an article with a minimum frequency of once per day and maximum of once per 30 days. The system is able to check changes in the HTML content for the specified URLs, if changes occur, the system will generate another time-stamp for this article. If the user provides an email address, during schedule time-stamping the system sends an email if changes occur in the web article. The system may also check online news articles in four different countries with a schedule. If the time-stamped article was modified in the selected country, the system would generate another time-stamp. In addition, the user is notified with an email.

### 3.4.7 Search and Browse articles

The user may search news articles by giving a word or URL or by browsing different domains. Domains are alphabetically arranged for ease. The searched articles are also shown with a calendar to help find user articles with appropriate dates of time-stamp. The calendar shows yearly, monthly, weekly and daily views. The search in the system searches saved URLs, web article titles and post titles for a better search.

### 3.4.8 Comparisons of Changes to Articles

Two saved time-stamped articles can be compared with each other. This allows users to identify the exact words or sentences that changed in the previous and the new versions of the web article. Moreover, a time-stamped article can be compared with the current online content of the same article.

### 3.4.9 Comparison of article content for different countries

The comparison is also possible with the same article from another country. In order to find out if a web article is showing different content in different countries of the world. Initially, Russia, China, UK and USA and default location of the server (in our case Germany) is supported.

### 3.4.10 Site Usage Statistics

Statistics that show the most-visited URLs in the system, with respect to their country of origin. The percentage of each individual URL in the total web articles in the system is visualized.

### 3.4.11 Check for Blocked Articles

The system should be able to check if a web article is blocked in some parts of the world. Moreover, the user should be able to see on a world map, where an article in the world is blocked.

# 4. Evaluation

There are no existing systems available for trusted time-stamping of online news articles and keeping track of modifications in the time-stamped content. Thus, proper evaluation of a new system for such purpose tends to be difficult, since there is nothing to compare against. Therefore, we start usability evaluation with the following aims.

## 4.1 Aim of Usability Evaluation

The aim of usability evaluation is to improve the usability and the user interface of the system [37]. The success of the product depends on the usability of the system, since if it is not easy to use, the system will lack user satisfaction. The chances will be high that the system is not going to be a success. Usability goals include efficiency, effectiveness and satisfaction [38]. The following subsections discuss different goals of usability evaluation.

### 4.1.1 Ease of use

The user must be able to use the system and perform all the functionalities of the system with ease and in a straight-forward way. A usability study was conducted to assess the use for all individual functions of the system, and improve ease of use. Thus, the aim was to develop an easy and simple interface, which resembles objects in the real world, so that the user has no complications in using them. The study participants were given some tasks to complete and were observed for usability improvements. The test must contain all areas of the system. The study participants should be able to finish the tasks. The aim was to find out find during how long it takes a user/study participant to complete a task. If a participant takes more time than normal, the system lacks effectiveness, and may result in unsatisfactory users.

### 4.1.2 Ease of learning or familiarizing with the system

The goal of usability is that the system should be easy to learn [39] [40]. There could be many features in the system, which may be new for the user. It is important that user feels comfortable with the system and can learn the system intuitively. Ease of learning may include 'error messages' or 'information messages', etc. in a helpful and intuitive way, so that the user may learn the system as fast as possible. During the usability testing, the time to complete a task was logged and a task was improved if study participant took a longer than expected period of time in learning them.

### 4.1.3  Satisfaction of the user with the entire experience

The user satisfaction is also an important aspect for consideration for a user to actually apply a system. The tasks in the system should be easy and effective. In order to find the satisfaction of the user, some general questions have to be asked from the participants of the usability study. See appendix page 69 for survey details and list of questions asked from usability study participants. Hence, improvement will be made in the system. The user/participant satisfaction survey would help us improve usability and enhance user satisfaction.

## 4.2  Evaluation Methodologies

There are different methodologies for usability evaluation. There are inspection methods for usability evaluation such as 'Heuristic evaluation' [41] and Cognitive walkthrough [42] which require hiring of usability experts to perform usability evaluation. On the other hand, 'Usability Testing' is a black-box method, which involves observing real time users while they use the system [37]. Hiring experts for usability evaluation of a system is very expensive and sometimes the results of inspection methods may generate false positives. In addition, 'Usability Testing' can be slow and time-consuming, since each individual study participant needs to be observed and to be given some system related tasks to solve. However, the user has to use the system and learn. It is important that the system is learnable, and is improved by perceiving real potential system users. Moreover, both techniques require informing the participants about the background of the system. In our case, the 'Usability Testing' approach was found more convenient in order to perform usability evaluation of the system. Moreover, the 'summative usability evaluation' approach has been used instead of 'formative evaluation', since usability testing is conducted at the end phase of the system design.

Usability testing is easy to perform and developers of the system may observe the users to identify the problems they faced. Usability testing also gives developers the opportunity to see how a user will react while using the system. User's emotions may say a lot about the user experience of the system. Furthermore, usability testing also gives us an idea about the number of steps a user requires to finish a task, or how accurately a user may use the system. These questions may only be answered by using the 'Usability Testing' approach. The entire user sessions may or the voice of the users may also be recorded to analyze the problems faced by the user. Users' feedback gives a sophisticated idea about the usability of the entire system and major usability problems may be identified.

## 4.3  Usability Testing

The goal of the 'Usability Testing' method is to find out the reaction of the real user at the time of interaction with the system [38]. Moreover, finding out 'how easy was it for the usability participants to accomplish tasks in the system.' This is accomplished by providing tasks to different study participants, ideally from the same domain in which the system would be used. The participants were observed during the evaluation sessions by an observer. The users were observed in a controlled environment to determine how well a user can use the system [43]. In our case, the system developer observed all the Usability Study participants. At the end of the sessions, study participants were given questions to answer. The data was collected during the 'Usability Testing' sessions and finally a report was created, which contains problems faced by different study participants together with a usability rating from the participants during the walkthrough. Finally, the system was redesigned to overcome the issues faced by the participants.

### 4.3.1  Evaluation Process

User study participants (8 participants) were given 12 tasks (with some tasks containing sub-tasks resulting in a total of 15 tasks) to perform (see 'Usability Study questionnaires' in appendix for more details on page 69). Study participants were observed during the walkthrough sessions. In the end, they were asked questions regarding the usability of the system as well as their overall satisfaction with the system. The answers to questionnaires submitted by each study participant were carefully analyzed and an evaluator noted the tasks that a participant was not able to finish or found difficult to perform and those tasks were categorized according to Table 4-3. Moreover, users were asked to define the tasks, which they found difficult to perform. The severities of the problems were divided into five categories. The format of the usability scale was taken from [44].

The scale described in Table 4-1 was used to perform the user study and categorize the severity of usability problems encountered in the tasks during the study. The severity of usability is categorized in four different categories and in order to report each usability problem each level of severity is assigned a number as shown in Table 4-1.

Table 4-1: Description Usability severity rating used in the Usability Testing

| ID | Severity | Description | Fix |
|----|----------|-------------|-----|
| 0 | None | Not a usability problem | no need to fix |
| 1 | Normal | Cosmetic problem | need not be fixed unless extra time is available on project |
| 2 | Minor | Usability Problem | fixing this should be given low priority |
| 3 | Major | Usability Problem | important to fix, so should be given high priority |
| 4 | Critical | Usability Catastrophe | imperative to fix this before product can be released |

Study participants were given tasks to perform and those tasks contained subsets of all the functionalities in the system. Table 4-2 gives an overview of different types of tasks study participants were given to perform.

Table 4-2: Description of the 'Topics' of given tasks to the user

| ID | Task Topics | Number of Tasks |
|----|-------------|-----------------|
| 1 | Account management actions | 3 |
| 2 | Stamping and stamp verification actions: | 3 |
| 3 | Article comparison actions | 5 |
| 4 | Remote operations actions | 3 |
| 5 | Application usage statistics | 1 |

The participants were carefully observed during the study testing sessions. The participant's screens were recorded for later inspections. The study participants were also instructed to think out loud, so that the observer may know what participants are thinking and what is creating problems for them. The participant's voices were also recorded for observing study participants' problems later on. If a participant was unable to finish a certain task, that task was assigned a severity rating according to the problems faced by study participants. The study participants were also asked to suggest usability improvements. The criteria according to which the usability problems were distinguished are as follows [45]:

Table 4-3: Criteria of rating usability problems faced by the study participants during usability study.

| ID | Severity | Description |
|---|---|---|
| 0 | Cosmetic | Participant's suggestions if they can improve usability |
| 1 | Normal | Irritant: The problem occurs only intermittently, can be circumvented easily, or is dependent on a standard that is outside the product's boundaries. Could also be a cosmetic problem? |
| 2 | Minor | Moderate: The study participant will be able to use the product in most cases, but will have to undertake some moderate effort in getting around the problem. |
| 3 | Major | Severe: The study participant will probably use or attempt to use the product here, but will be severely limited in his or her ability to do so. |
| 4 | Critical | Unusable: The study participant is not able to or will not want to use a particular part of the product because of the way that the product has been designed and implemented. |

After each usability test, the usability fixes were done and tested with the second participant if she/he repeats the same issues again.

### 4.3.2 User study participants and Demographics

Jacob Nelson, a web usability consultant, proposed in 1993 that optimum results come from usability testing with five users and using as small tasks as one could afford [46]. The claim of five users is also described in a mathematical model in [47] using the following equation:

$$N = (1 - (1 - L)^n)$$

Here 'N' is the number of usability problems, 'n' is the number of users and 'L' is the probability of one user identifying a usability problem. When 'L' is used 31% in a research by 'Tom Landauer' [48] the graph in Figure 4-1was obtained.

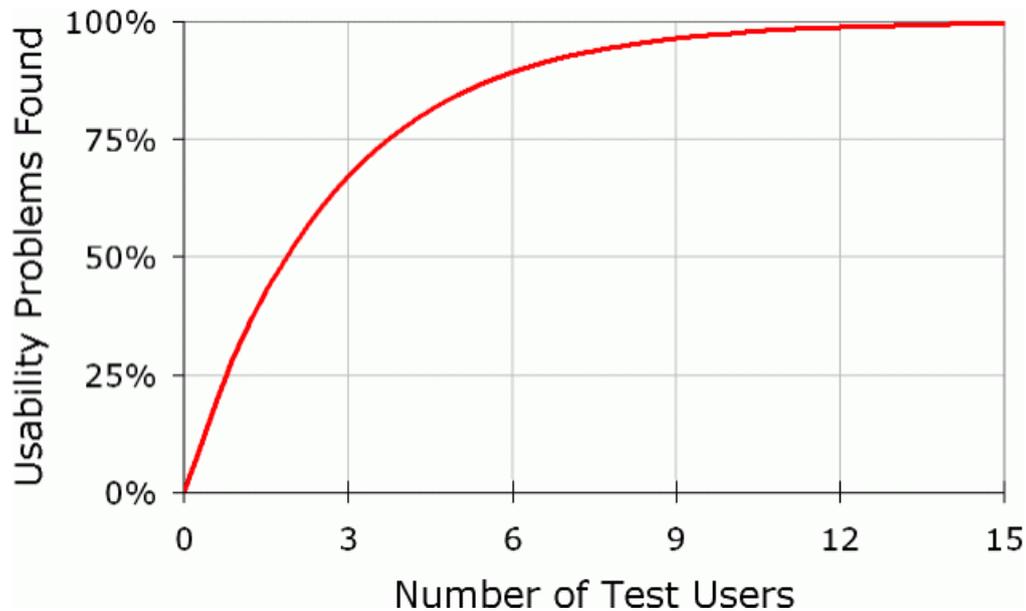

Figure 4-1: Number of usability problems versus the number of Usability Test users. [48]

The graph shows that if only five users are used for 'Usability Testing' about 80% of the usability problems have been eliminated. In order to get maximum benefits from users and detect as much usability issues as possible it is ideal to use five to ten users. Thus, following this strategy, we used eight users for our 'Usability Testing' study. The user of the system might be normal people from all occupations. Hence, students were recruited for usability testing sessions from different disciplines and different age groups. One student was also a PhD candidate for economics and politics, since the main aim of this system might be for people from politics and journalism. Other than that, participants from different age groups and different fields were recruited. Participants were asked to use the system and they should think out loud so that the evaluator may know what they are looking for and what they find problematic in the system. Table 4-4 describes further details of the usability testing participants.

Table 4-4: Background and Demographics of user study participants

| ID | Background | Gender | Age |
|----|------------|--------|-----|
| 1 | Researcher (science) | Female | 32 |
| 2 | Master Student Computer science | Male | 28 |
| 3 | PhD Candidate Economics and Politics | Female | 30 |
| 4 | Master student Sports sciences | Male | 27 |
| 5 | PhD Candidate Physics | Male | 24 |
| 6 | PhD Candidate Economics | Male | 30 |
| 7 | Bachelors Student Process and Environmental Engineering | Male | 26 |
| 8 | Bachelors Student Architecture | Male | 21 |

## 4.4  Evaluation Results

There were eight evaluations sessions and after each session some minor usability fixes and cosmetic fixes were done. However, if usability fixes would have to take a long time they were fixed after the sixth usability session. The last two study participants were able to finish all the tasks without any difficulty. On average, each usability session took about 40 minutes. There were a total of 12 tasks for each participant. Tasks were also changed and modified for some participants, or some tasks were divided into sub-tasks, which were confusing for the study participants to complete at once.

Table 4-5: Table shows average severity rating of all the tasks given to the study participants as well as the average severity rating for all the participants

| ID | Tasks | Participant 1 | Participant 2 | Participant 3 | Participant 4 | Participant 5 | Participant 6 | Participant 7 | Participant 8 | Average Severity Rating |
|---|---|---|---|---|---|---|---|---|---|---|
| 1 | Task1 | | | | | | | | | 0 |
| 2 | Task2 | 2 | 2 | | | | | | | 0.5 |
| 3 | Task3 | | | | | | | | | 0 |
| 4 | Task4 | | | | | | | | | 0 |
| 5 | Task5 | | | | | | | | | 0 |
| 6 | Task6 | | | 3 | 2 | | | 1 | | 0.75 |
| 7 | Task7 | 3 | 2 | 3 | 2 | 3 | 3 | | | 2 |
| 8 | Task8 | 3 | | 1 | 1 | | | | | 0.625 |
| 9 | Task9 | | | | | 3 | | | | 0.375 |
| 10 | Task 10 | | 1 | | | | | | | 0.125 |
| 11 | Task 11 | 1 | | | | | | | | 0.125 |
| 12 | Task 12 | 2 | 2 | | 1 | 3 | 2 | 2 | | 1.5 |
| | | | | | | | | | | |
| | Total Tasks with usability Problems | 5 | 4 | 3 | 4 | 3 | 2 | 2 | 0 | |

As we can see from the table, the total number of usability problems identified was incrementally decreased throughout the cyclical evaluation due to improvements in usability of the system after each usability testing session. Figure 4-2 shows the average user feedback after the usability testing where '9' is the highest value and '0' is the lowest. We can assume that the system has significant 'ease of use' and is user friendly. Moreover, Figure 4-2 describes some insights of overall user satisfaction after using the system.

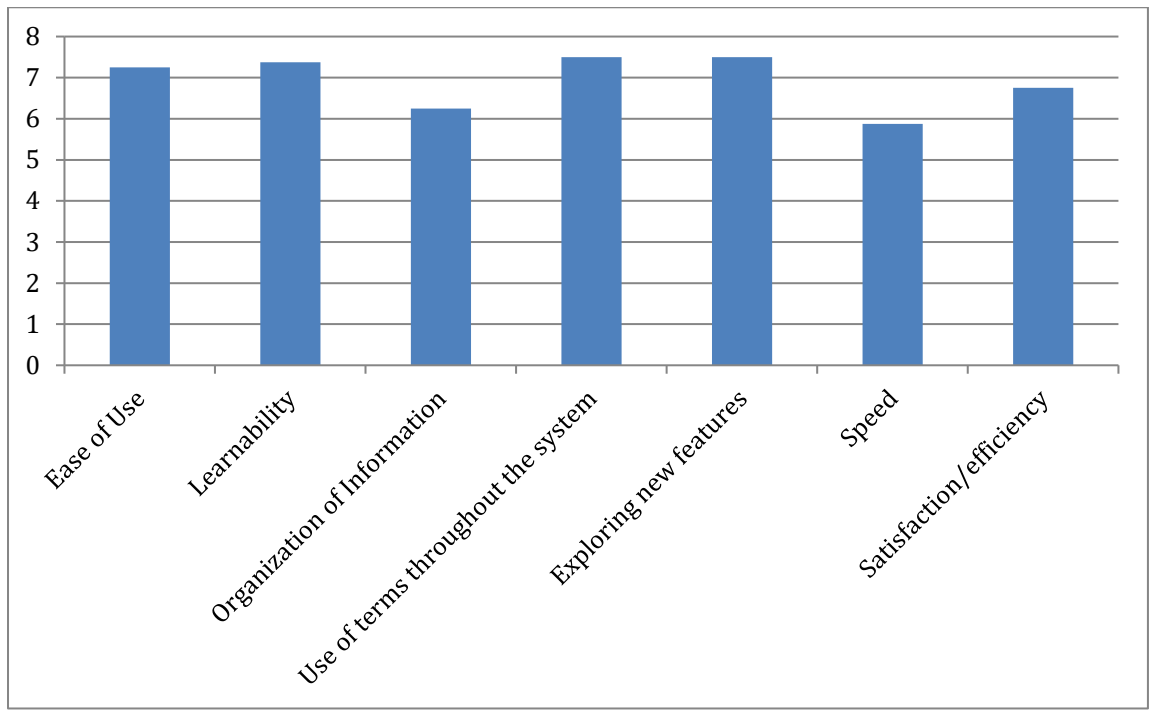

Figure 4-2: Ratings for the usability testing by the users. '0' as the lowest score and '9' is the highest.

## 4.5 Discussions

If we analyze the results from Table 4-5 most of the study participants found problems in 'task 7'. One reason for this could be that this task actually consists of three subtasks and most of the participants were able finish two subtasks. If a participant completed two out of three subtasks, then it was nonetheless considered incomplete. Thus, we found more severity ratings in 'task 7'. The system provides different comparison options and participants were unable to distinguish between them and started believing that they already finished some subtasks. However, with time, users may find the system easy and may differentiate different comparison options. After changes to the design of 'comparison options tasks', we find significant change in usability. Thus, three separate buttons were provided for comparing articles with 'Current online version' with 'Another saved time-stamp' and with 'Articles from another country'. This change might improve the usability and users may understand the difference between the three different options.

Task number 12 also has a high severity rating of 1.5. Most of the participants were unable to understand the visualization in task 12. Improvements in the legend of the visualization were made and decimal values were reduced up to '2' decimal places for better understanding.

'Task 2' is simple email confirmation when a user receives an email for registration confirmation and clicks the provided link and everything is done successfully. Initially, the system did not redirect to the user confirmation page if a user was not logged in to the system. Hence, some participants find it unclear that they already confirmed their email and system is still prompting to confirm their email. Thus, a 'confirmation page' was redirected after the login page. After this step, all other participants finished this task smoothly.

'Task 6' was exactly the same as 'task 5', except that the participants need to select the country in 'task 6' to compare the article with. Thus, some participants started searching 'task 6' in the window of 'task 5'. Similarly, 'task 8' was also difficult to find by one participant and if at first, the user first explored the whole system and has an idea about the system as a whole, then the user would be able to perform tasks more easily. In order to help users with the different functionalities of the system, a 'FAQ' page was added in the system which lists all the different 'core options' the system performs. Furthermore, many cosmetic changes have been performed in order to improve readability and ease of use. For example, associating 'glyph-icons' with each option and adding text to clarify to the user what exactly an option does. However, the feedback from the users were very positive as can be seen from the results of usability ratings  questionnaires in Figure 4-3.

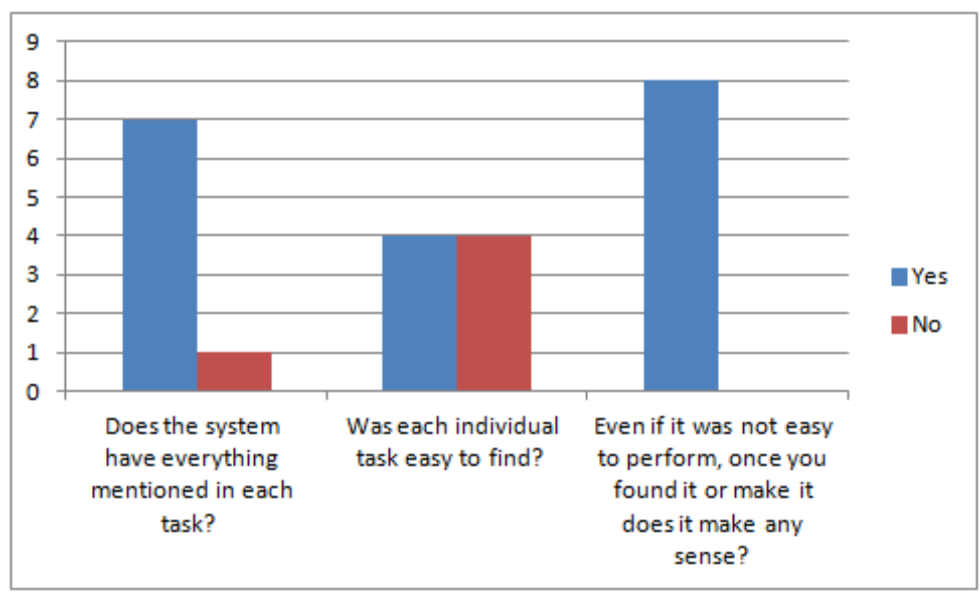

Figure 4-3: Usability testing feedback from the users

# 5. Outlook

## 5.1 System Improvements

This section describes additional features and improvements that could be integrated into 'StampTheWeb' in the future.

### 5.1.1 Articles by modifications

The system may include statistics for articles which have highest modification, once time-stamped in the system. Then the user could discover which domains or media outlets tend to modify their reporting and maybe also modify their political opinions.

### 5.1.2 Help

Help and reporting bugs sections are not included in the system. It is necessary that the user is able to find some components in the system. A Help section may also contain videos that visualize performing certain tasks in the system, this can also make the system very user friendly.

### 5.1.3 Support for IPFS

Currently, the system supports HTTP and HTTPS. It would be a very good idea to take the system to the distributed web technologies like IPFS. IPFS will make it faster due to the distributed access of files from different locations. Moreover, it is more secure since files are not present in a centralized location.

### 5.1.4 Check multiple proxies

In order to find if a web page is blocked in a country, we only check one proxy. If that proxy is unable to find the requested page, it is deemed blocked in that country. However, sometimes a proxy doesn't work and in that case, deeming it blocked is false. We need to add more than one proxy, at least three per country.

### 5.1.5 Block map for Entire world

Currently, the system checks for block map for 31 different countries of the world. We have map coordinates for the entire world. We need to add support for all the countries of the world, so that the user may find where in the whole world an article is blocked.

### 5.1.6 Improved Search

A user is able to see web articles time-stamped by any user. A user may be interested in time-stamped content from different users related to different subjects. By using the advanced search, we may give users an opportunity to find the content of interest in a sophisticated way. For example, the search should not only show the 'searched keyword' results, but also related results. Databases

Currently, the database supports a maximum of two terabytes of data. If the size of the database exceeds this limit, then we need to migrate the database to other databases like MySQL or NoSQL databases.

### 5.1.7  Login using other accounts

Currently, the user is required to create a new account in order to start using 'StampTheWeb'. It just takes a few minutes to create and activate an account using a personal email account. However, nowadays, it is getting difficult to create so many different accounts online, yet, it could be easier for a user to login using another system, such as social media domains like Facebook, Twitter and WordPress.

### 5.1.8  Sub domains

Currently, while time-stamping, the system extracts text of the single provided web-page and only considers the content of that particular page. However, the system should also check for sub-domains present in the provided URL. Moreover, the user should be given different options for extracting the text from only the provided URL or going through one or two sub-levels present in the provided URL.

### 5.1.9  Time-stamp Reference Plug-in

If some online web sources are referencing other web sources. They can time-stamp the referenced sources into our system and request us time-stamp plug-ins for the referenced content. This way they can link time-stamps of referenced sources into their web-pages. An example of this would be a Facebook share button by using which anyone can share their personal online news articles in their system. By having a time-stamped plug-in the trust of users on the website would increase since the page which is been referenced existed at certain time and is a trusted time-stamp which can be verified by clicking the button or visiting the original time-stamped page at 'StampTheWeb'.

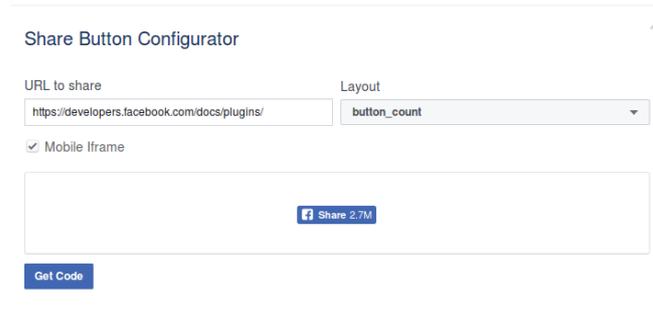

Figure 5-1: Facebook share button can be added to any web-page and can be used to share a personal webpage into Facebook. [49]

Some usability improvements have been added in Appendix section 8.2

# 6. Conclusion

A free to use time-stamping service for online news articles has been introduced. Time-stamped articles are secure and verifiable due to the use of Bitcoin blockchain. In addition to time-stamping news articles, which was the specific problem setting addressed in this thesis project, any other web articles can also be time-stamped, for example, Wikipedia pages, Blog posts, etc.

Automatic time-stamping is possible with the given system for news articles by providing a certain frequency. Moreover, user may get notified by email address, if information changes in the time-stamped article. This system uses secure technologies for saving the results of the time-stamped content for example embedding hashes into Bitcoin blockchain (which is cryptographically validated blockchain). Thus, information is present in a distributed blockchain network, making it impossible to get manipulated. Due to the fact that the creator of the currency states it is impossible to manipulate information present in the Bitcoin blockchain unless attacking nodes exceed more than 50% [33]. The user is not supposed to rely on a centralized time-stamping authority for time-stamp creation with this service. Furthermore, users of the system have the following advantages, presented in the thesis:

- Find out if a news article has been changed at any point in time compared to its first time-stamped version.

- Get notified for changes in a news article by email or present the changes or prove changes were made by news sources.

- Get informed about news blockage in a country or anywhere in the world.

- This system is the first initiative in time-stamping online information or news articles. Existing systems only time-stamp offline digital information.

- There are very few systems, which may record changes in the news content. However, they do not possess any verifiability of their saved content.

- Easy to use with different views that help the user perform different tasks conveniently.

- All the information present in the system is public and anyone can get benefit from it.

The system presented in the context of this thesis contributes to the field of journalism and supports the user in quickly identifying changes made to news articles that would previously have remained unidentified.

# 8. Appendix

## 8.1 Usability Study Questionnaires

# Questionnaires

**Personal Details**

Name:

Profession:

Age:

Field of Work / Study (e.g. Scientist / Software Engineer):

**Dear Participant**

We welcome you and appreciate your participation in the following questionnaire about 'StampTheWeb. Below is a short description of 'StampTheWeb to assist you:

**What is 'StampTheWeb?**

Stamp The Web is a trusted time-stamping service for web-based content that can be used free of charge. The system can be accessed using URL https://stamptheweb.org. The service enables you to automatically create trusted time-stamps to preserve the existence of online content at a certain time point, such as newspapers articles, blog posts, etc.

**Why Create Time-stamp of online Content?**

This enables you to certify that certain information existed online in a particular state at the time it was 'trusted time-stamped' using 'StampTheWeb'.

**What is a Trusted Time-stamp?**

The information is hashed and sends to the Bitcoin network, and later it can be proved from Bitcoin's network. Bitcoin is relying on a peer to peer network and does not possess a centralized server and it is proven to be very secure so far. Changing information stored in the Bitcoin transaction history is not possible. Thus, it is a trusted time-stamp and eventually you have a proof for an online resource. If any change in the online information were made, the created hash would be different. To find out more about the technical details of this system please visit:

http://stamptheweb.org

http://www.originstamp.org/

https://en.wikipedia.org/wiki/Bitcoin_network

**Things you can do with 'StampTheWeb**

1) Stamps a web article, news article or blog post, etc.
2) Compare two saved (time-stamped) web pages with each other or with the current online version.
3) Search for other saved articles by text or by categories.
4) Compare a time-stamped article with the same article from any other part of the world for e.g. Russia, UK, USA, etc.
5) Compare a time-stamped article from a certain date with its current online version.
6) Compare a time-stamped article from a certain date with its time-stamped version from another date.
7) Check if an online article is blocked in any other part of the world.
8) Check where in the world an article is blocked through a map.
9) Review statistics of web pages most visited in the system with respect to country of origin.
10) Time-stamp an article according to a user-defined schedule. If any change is detected in that article the user is notified by email (if email is provided), or information is automatically stored in the system.
11) Time-stamp an article according to a user-defined schedule and compare it at regular intervals with the same page from other countries. The user is notified by email if changes are found.
12) Account actions: Manage your account, passwords, email, profile, etc.

**Your task is to simply use StamptheWeb and we will find out how easy was it for you to use.**

**Account management actions:**

Task 1 Create your profile

Task 2 Receive email for new registration and confirm your account

Task 3 Login into the system

**Stamping and stamp verification actions:**

Task 4 Stamp a web page (any web page of your choice news article, Wikipedia, etc.)

Task 5 Verify the creation of a trusted time-stamp...

Task 6 Make a scheduled time-stamp for a web page of your own choice

**Article comparison actions:**

Task 7 a) Compare a time-stamped web page with the current online news articles of that web                page.

b) Find all the posts from one domain for example (wikipedia.org, or New York Times)

c) Compare two already saved time-stamped web pages with each other. Both web pages should be from the same domain for example (wikipedia.org, or New York Times) and time-stamped on different months for example (May and June)

Task 8 Compare a Web page with the same web page in another country

**Remote operations actions:**

Task 9 Make a schedule time-stamp for a web page of your own choice and then compare it with the same page in another country. Also, provide your email so that you are notified in case of changes in the page.

Task 10 Check if a web page is blocked in a country like China, Russia, USA or UK.

Task 11 Check where in the world an article is blocked.

**Application usage statistics:**

Task 12 Check statistics. What do you understand from it? Tell us verbally.

**Note:**

- Try to complete the tasks as if you were doing this for real. Spend as little or as much time as you normally would do these tasks.
- It is not compulsory to complete all the tasks.
- It is not your testing. It is the testing of the system.
- If you are not able to complete a task, it helps us improve the ease and usability of that task.
- The computer screen will be recorded in order to improve the system.
- Please answer the following questions if possible.
- Please think out loud i.e. your steps, what are you looking for, etc. Therefore, we can understand where you are getting problems.

**Questions**

1) Does the system have everything mentioned in each task?
a) Yes

b) No
c) What is missing?

2) Was each individual task easy to find?

a) Yes
b) No
c) If no then Task no:

3) Which task/ tasks took more time than usual?

a) Tasks:
b) Any Improvements:

4) Which task/ tasks were difficult ones?

a) Tasks:
b) What was difficult?

5) Any other relevant suggestions to make it more user-friendly?

a) Suggestions (if any):

6) Even if it was not easy to perform. Once you found it or make it, does it make any sense?

a) Yes
b) No

**User Satisfaction**

7) Overall satisfaction of use for the whole system in scale of '0' to '9' where '9' is the ideal.

|  |  | 0 | 1 | 2 | 3 | 4 | 5 | 6 | 7 | 8 | 9 |  | N/A |
|---|---|---|---|---|---|---|---|---|---|---|---|---|---|
| Ease of Use | Easy |  |  |  |  |  |  |  |  |  |  | Difficult |  |
| Learning to operate the system | Difficult |  |  |  |  |  |  |  |  |  |  | Easy |  |
| Organization of information | Confusing |  |  |  |  |  |  |  |  |  |  | Very Clear |  |
| Use of terms throughout system | Inconsistent |  |  |  |  |  |  |  |  |  |  | consistent |  |
| Exploring new features by trial and error | difficult |  |  |  |  |  |  |  |  |  |  | Easy |  |
| System speed | Too slow |  |  |  |  |  |  |  |  |  |  | fast |  |
| Satisfaction/ Efficiency | unsatisfied |  |  |  |  |  |  |  |  |  |  | satisfied |  |

## 8.2 Usability and System Improvements 'StampTheWeb'

### 8.2.1 Different views for Time-stamp Posts

Currently system displays time-stamps as 'posts in a blog' and it is better to provide a detailed view [50] if large amounts of posts are searched. Detailed view helps users find insights about the selected content. Moreover, a thumbnail view [51] would help identify time-stamps by images and better visualization.

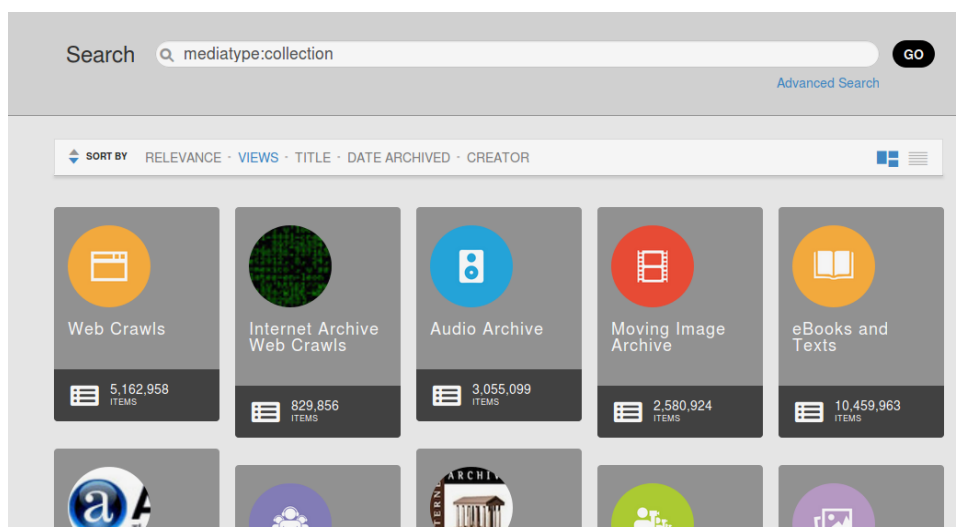

Figure 8-1: Detail view of showing files and folders in Windows 10.

Figure 8-2: Thumbnail view of digital content in archive.org

### 8.2.2 Improved Search for time-stamped articles

Currently, the search functionality can be used to search the time-stamped articles. However, there is no way to search for other pages like 'Block Articles'. It is not user friendly to type the search keyword and press the button in order to find the requested content. The content should be available to the user as soon as he or she types the keyword. The example would be a Google search, which consults the user without page load.

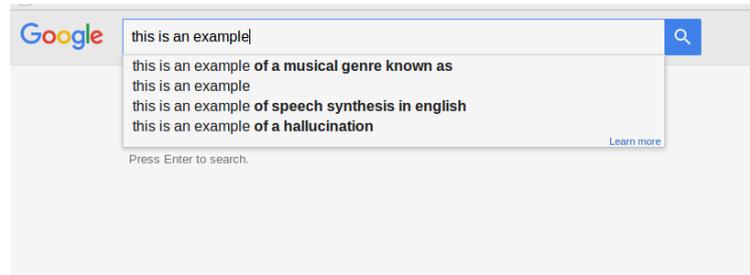

Figure 8-3: An example of Google search and user is being suggested without the page being loaded.

### 8.2.3  Deleting/De-activating User Profiles

A user may register in the system, but there is no option to de-activate the user profiles, yet this is an important issue and the storing of data without the user's consent should not be possible.

### 8.2.4  Showing different versions of same article

Currently a user may browse articles with the given 'keywords' or find articles from the same domain. The user may also get similar articles by using the 'calendar view'. However, finding one article with its various versions or time-stamps with respect to different dates either in case of automatic time-stamp generation or by user modification is not possible in the current system. Showing different versions of same articles together will be an important feature since, the user may decide which two dates in order to compare an article.

## 8.3 Technical Documentation of 'StampTheWeb'

# Table of Contents

# 1 Introduction

This document describes the 'StampTheWeb system in detail. Besides discussing functionalities, it highlights the technologies, modules and APIs used in the development of the system.

## 1.1 Purpose

This document is designed to help understand the technical aspects of the system. With the help of this document, any software developer can enhance the functionality of this system, or make improvements in the future, if required. This document provides sufficient information for a programmer to produce the software, and for a maintainer, who may not be the developer himself, to make subsequent changes.

## 1.2 Scope

'StampTheWeb' allows its users to partake in securing digital heritage online. 'StampTheWeb' is a trusted time-stamping service for web-based content that can be used free of charge. The service enables users to automatically create trusted time-stamps to preserve the existence of online content at a certain point in time, such as newspapers articles, Wikipedia pages, blog posts, etc. This enables users to prove that certain information online existed in a particular state at the time it was 'trusted time-stamped' using 'StampTheWeb'. It can be accessed online through "https://www.stamptheweb.org". In this documentation 'this' or 'the' 'system' are both abbreviations for 'StampTheWeb.

### 1.2.1 Developing a tool for archiving and comparing news articles on the Web.

For scientists and information researchers we require a tool for efficient access to the web archived content[1]. Ease of use, extensibility and re-usability is also important for a web archive system. The system should be fast and easy to use at the same time in order to encourage usability.

'StampTheWeb stores the hashes, or essentially the time-stamps, of the online content independently. These hashes are sent to Bitcoin's blockchain with the help of the 'OriginStamp' API [2] for later verification from Bitcoin's decentralized network. Thus, the user has the opportunity to verify time-stamps of any online content from 'OriginStamp' as well as from the Bitcoin network. 'StampTheWeb also redirects to 'OriginStamp' for verification. All time-stamped content is associated with a hash value, which represents the online information that is supposed to be time-stamped. This hash value is required in order to verify the time-stamps from 'OriginStamp. Some of the functionalities of this system are as follows:

- Stamp a web article, news article or blog post, etc.
- Compare two saved (time-stamped) web pages with each other or with the current online version.
- Search for other saved articles by text or by categories.
- Compare a time-stamped article with the same article from any other part of the world e.g. Russia, UK, USA, etc.
- Compare a time-stamped article from a certain date with its current online version.
- Compare a time-stamped article from a certain date with its time-stamped version from another date.
- Check if an online article is blocked in any other part of the world.
- Check where in the world an article is blocked through a map.
- Review statistics of web pages most visited in the system with respect to country of origin.
- Time-stamp an article according to a user-defined schedule. If any change is detected in that article, the user is notified by email (if email is provided), or information is automatically stored in the system.
- Time-stamp an article according to a user-defined schedule and compare it at regular intervals with the same page from other countries (selected countries). The user is notified by email if changes are found.
- Account actions: Manage your account, passwords, email, profile, etc.

---

[1] "Proceedings of the 16th ACM/IEEE-CS on Joint Conference on Digital ..." 2016. 23 Jul. 2016 <http://dl.acm.org/citation.cfm?id=2910896&picked=prox>
[2] "OriginStamp: Trusted Timestamping with Bitcoin." 2016. 23 Jul. 2016 <https://www.originstamp.org/>

### 1.2.2 Showing similar records together

'Corpora building' is a fundamental task for a Web Archive system[3.] It means showing similar posts, results together in a meaningful way. We show similar results from posts together, when a keyword is searched. Same categories are kept together in 'browse other submissions'. We make categories by using domain names. In order to efficiently use saved information, we are using a temporal view whenever information is searched for or browsed through. We provide easy access of archived data to political research scientists by dividing content into categories and showing timelines with different views.

---

[3] Holzmann, Helge, Vinay Goel, and Avishek Anand. "ArchiveSpark: Efficient Web Archive Access, Extraction and Derivation." *Proceedings of the 16th ACM/IEEE-CS on Joint Conference on Digital Libraries* 19 Jun. 2016: 83-92.

'StampTheWeb serves many purposes; it can for example serve as a web archive, online heritage protection system or as a web articles aggregator. There are many existing applications for similar purposes (i.e. web heritage protection) such as archive.org[4], Google news archive[5], CNN news archive[6], BBC news archive[7], etc. While these sources can somehow protect information heritage on the web, they are not secure, because information could be hacked, manipulated by administrators, or there is a possibility of server failure. This system protects online web heritage, as well as offers the tamperproof time-stamping of online content, in particular, online news coverage or other information of the user's interest. In the 'Internet Archive' one can archive any web content for the future to see, but it does not add information to the Bitcoin network to enable the verification of the content. In contrast with 'StampTheWeb, our aim is to provide a secure web archive with the possibility to verify the information. Thus, 'StampTheWeb presents a system with strategically advantageous and better security mechanisms in terms of verification and transparency.

## 1.3 Definitions, Acronyms and Abbreviations

**WSGI**: The **Web Server Gateway Interface** (**WSGI**) is a specification for a simple and a universal interface between web servers and web applications or frameworks for Python. [8]

**Satoshi**: A Satoshi is the smallest fraction of a Bitcoin that can currently be sent: 0.00000001 BTC, that is, a hundredth of a millionth BTC.[9]

**Flask Blueprint**: A blueprint defines a collection of views, templates, static files and other elements that can be applied to an application.[10]

**Jinja2:** A templating language for HTML documents, to enable inheritance in HTML documents and thus increases the speed of development and reduces redundancy in HTML.

---

[4] "Internet Archive: Digital Library of Free Books, Movies, Music ..." 2006. 23 Jul. 2016 <https://www.archive.org/>

[5] "Browse all newspapers - Google News." 2011. 23 Jul. 2016 <https://news.google.com/newspapers>

[6] "CNN Student News - Archive - CNN.com." 2015. 23 Jul. 2016 <http://www.cnn.com/specials/student-news-transcripts>

[7] "BBC Archive." 2007. 23 Jul. 2016 <http://www.bbc.co.uk/archive/>

[8] "Web Server Gateway Interface - Wikipedia, the free encyclopedia." 2011. 23 Jul. 2016 <https://en.wikipedia.org/wiki/Web_Server_Gateway_Interface>

[9] "terminology - What is a 'Satoshi'? - Bitcoin Stack Exchange." 2011. 23 Jul. 2016 <http://bitcoin.stackexchange.com/questions/114/what-is-a-satoshi>

[10] "Blueprints — Explore Flask 1.0 documentation." 2016. 23 Jul. 2016 <http://exploreflask.com/en/latest/blueprints.html>

## 2 System Overview

The system has been designed using Flask. Flask is a new python web framework in comparison to other python web frameworks like Django, Pyramid, etc. As Flask is a 'micro web application framework' and does not require large frameworks in comparison to Django or Pyramid, it has the tremendous ability to help develop applications faster. Even so, Flask applications are extensible[11]. Flask uses Jinja2, which is a templating language for python. Thus working on the front end of 'StampTheWeb requires basic understanding of Jinja2. Flask is based on 'Werkzeug', which is a Web Server Gateway Interface (WSGI) library for python.

Due to the above-mentioned advantages, Flask is found to be suitable for creating this application. Flask has different packages that support sophisticated applications development. Besides using Flask, the application is tested in python 3.4.2 and upwards as well as for python version 3.5.2. On the deployment server on Amazon EC2, we are using python 3.4.3. Python 3 was the logical choice as python 2.7 support is only available for four more years, since 'end of life' for python 2 is in 2020. Hence using python 3 is a good option for the future.

Python has tremendous support for remote operations, such as accessing web pages from different remote locations, for example through proxies, and therefore it also was a good option for 'StampTheWeb. For instance, the proxy features of python help us find out if a web page is blocked in another location in the world and they help us find out locations of web articles visited. Only free python libraries have been used in the entire project. For better understanding of the python packages used, all the different packages used for specific operations are discussed later in chapter 5.

## 2.1 System Characteristics

'StampTheWeb uses secure HTTP i.e. HTTPS to be accessed through the web. It is attempted to use global variables throughout the system for different configurations. If these variables are stored into the operating system, the configuration of the system becomes easy. Thus, it is easy to append new features into the system. It just requires changing global variables or adding new global variables. For example, switching to another database requires a change in the following environment variable:

```
DEV_DATABASE_URL = <url for the DB>
```

---

The system first reads a variable from the environment variables of the operating system if available; otherwise, it takes default values from the code. For good practice, environment variables should be configured. The following commands add global variables to the 'Heroku' web server:

```
$ heroku config:set MAIL_USERNAME=<your-gmail-username>
$ heroku config:set MAIL_PASSWORD=<your-gmail-password>
```

In the server setup, the command is the following:

```
$ MAIL_USERNAME=<your-gmail-username>
$ MAIL_PASSWORD=<your-gmail-password>
```

Alternatively, the variables can be set in ~/.bashrc directly.

The provided email account through these variables would be used to send emails from the system, for example new registration, account confirmation, etc. Furthermore, different types of configurations are available depending upon which environment the system has been started with, i.e. Development, Testing or Production.

### 2.1.1 Login and Security

The Flask-Login module has been used to make login sessions secure. Using 'strong' protection, the application automatically logs users out, if the IP-address of the user has changed. Furthermore, a token based authentication scheme has been used with a very short expiration time (i.e. 3600 seconds) for each token. The client must send the authentication token with each request to the server in order to authenticate himself. As soon as a token expires, the client must re-authenticate to the system to get a new token. Flask's module 'itsdangerous' has been used to generate and validate the tokens. A user must provide an email in order to register. The user receives an email in order to activate the account and in order to start using the system. A strong password containing at least six characters is required for creating an account.

## 2.2 System Architecture

The system does not use any specific architecture, however, in order to efficiently manage application code, different blueprints have been used. The system's front end files have been separated from script and backend files. We are using Flask and it is a micro web framework. However, different modules have been separated in order to support organization of the code. Flask also supports blueprints and so two blueprints have been used. Flask blueprints are ideal for larger applications, in order to factor applications into a set of blueprints.[12] Thus, figure 1 depicts the component view of the system based on the two blueprints (main_blueprint, auth_blue_print). If a request arrives, related to authentication such as login, registration, authentication, it is handled using the 'auth_blueprint'. Other requests are handled by the 'main_blueprint'.

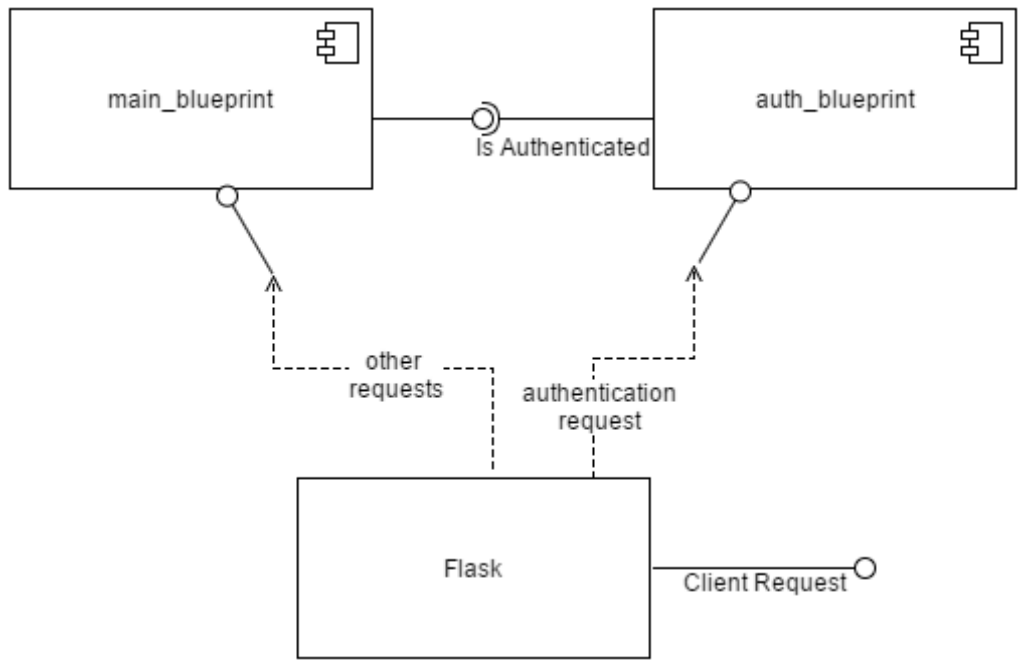

Figure 1: Component diagram 'StampTheWeb

Both these blueprints contain different set of files, folders and other blueprint specific documents.

## 2.3 Infrastructure Services

Utmost care has been taken during code-writing so that whenever an input or output operation is performed, the application should be able to handle an exception that occurs during the process. The error messages and information messages are reported in a sophisticated way using 'Flash' error messages by Flask. Thus, if an unwanted error occurs the user should be able to report, or at least know what is going on. A sample 'Flash message' is shown in figure 2. One example code for showing Flash message to the user as well as handling exceptions is given below:

```
try:
    pdfkit.from_url(url, path)
except OSError as e:
```

---

[12] "Modular Applications with Blueprints — Flask Documentation (0.11)." 2014. 23 Jul. 2016 <http://flask.pocoo.org/docs/latest/blueprints/>

```
if not app.config["production"]:
    flash(u'Could not create PDF from ' + url, 'error')
```

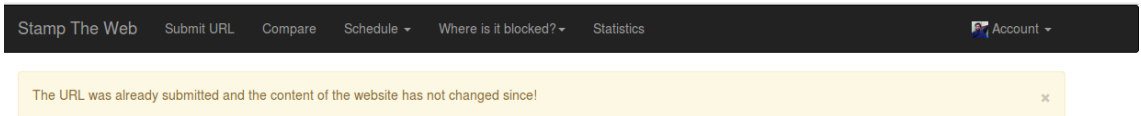

Figure 2: Flash message in 'StampTheWeb.

Similarly, in order to support developers and testers, background logs and information are collected. If any error or any other activity occurs, it is logged into the 'webStamps.log' file, which is present at the root directory of the source code. Not only error messages, but also information messages, are recorded throughout the system. An example from 'webStamps.log' file is shown below, where it indicates that a new PDF file with the given name has been created into the system. Later, a new Post has been added into the system with the date and time at which this was done.

'webStamps.log'
However, local PDF exists at
app/pdf/QmZsoGMPq7Jr1gVtJzojqaR6gkKmRMSmTKQysF1epiQYDJ.pdf
2016-07-27 17:25:12 New Post added

It is a good practice to report errors and other changes happening in the system into logs. The sample code for logging errors and information into the logs is as follows:

```
current_app.logger.info(datetime.now().strftime("%Y-%m-%d %H:%M:%S") +
" New Regular task added")
current_app.logger.error(datetime.now().strftime("%Y-%m-%d %H:%M:%S") +
" An error occur")
```

The object 'current_app' could be used to get the flask 'app' and app has already started a 'logging handler' into the 'init.py' file, at the time of application creation. Thus we can simply use 'logger' anywhere in the code.

# 3. System Context

'StampTheWeb uses 'OriginStamp' API[13] to generate time-stamps from the hashes that correspond to online content. We consider only text from the provided web articles and exclude pictures, ads, links, etc. 'OriginStamp' which uses aggregated hashes once a day for Bitcoin's transactions and only generates minimal Bitcoin transaction fees, i.e. 1 Satoshi. In return 'OriginStamp' API provides us the time-stamp against the provided hash. Although the hash is still not sent to Bitcoin blockchain. Hence, at this point the hash will not be verifiable in the Bitcoin network yet. It requires one day for 'OriginStamp' to collect all the hashes and send them to the Bitcoin blockchain.

Bitcoin is a widely used crypto-currency and, so far, it is found to be secure. Bitcoin was introduced in 2008 [14] and it has the highest participating nodes on the internet. The larger number of nodes is considered a security feature as discussed by the creator of this system. It is impossible to generate a Bitcoin attack unless the attacking nodes exceed 50% [33]. Furthermore, Bitcoin also uses a decentralized blockchain to store transaction information.

This transaction information is tamperproof due to the use of blockchains from the Bitcoin network. Furthermore, tampering would require a change in the whole transaction tree (based on the idea proposed by Merkle in 1988 [15]) as the most recent transaction contains the hash of all previous transactions. In order to make this service tamperproof, Bitcoin blockchain has a sophisticated method of storing this information as a transaction by forming a hierarchical architecture by using Merkle tree and nonce in such a way that changing any information would be quite impossible. In addition, Bitcoin stores information of all the parent transactions providing it with another layer of security, and tampering with any information needs a change in all upper levels of the transaction tree. For more information on 'OriginStamp' and Bitcoin blockchain please see review. [16]

'StampTheWeb simply sends hashes of the requested web pages to 'OriginStamp' using python requests. In the reply, 'OriginStamp' sends the time-stamp and related information using JSON format. The sample code for generating time-stamps is shown below:

```
headers = {'Content-Type': 'application/json', 'Authorization': 'Token
token="<Token To Be Generated from API"'}
data = {'hash': hash_variable, 'title': title_of_web_page}
requests.post(api_url, json=data, headers=headers)
```

---

[13] "Developer resources - OriginStamp." 2016. 6 Aug. 2016
<https://www.originstamp.org/developer>
[14] Nakamoto, S. "Bitcoin: A Peer-to-Peer Electronic Cash System." 2014.
<https://bitcoin.org/bitcoin.pdf>
[15] Merkle, Ralph C. "Fast software encryption functions." *Conference on the Theory and Application of Cryptography* 11 Aug. 1990: 477-501.
[16] Gipp, Bela, Norman Meuschke, and André Gernandt. "Decentralized Trusted Timestamping using the Crypto Currency Bitcoin." *arXiv preprint arXiv:1502.04015* (2015).

In response to the above code, a JSON value is returned and subsequently that 'Date' field contains the time-stamp information of the provided hash.

Figure 3 shows the deployment view of 'StampTheWeb and describes how operates with other systems.

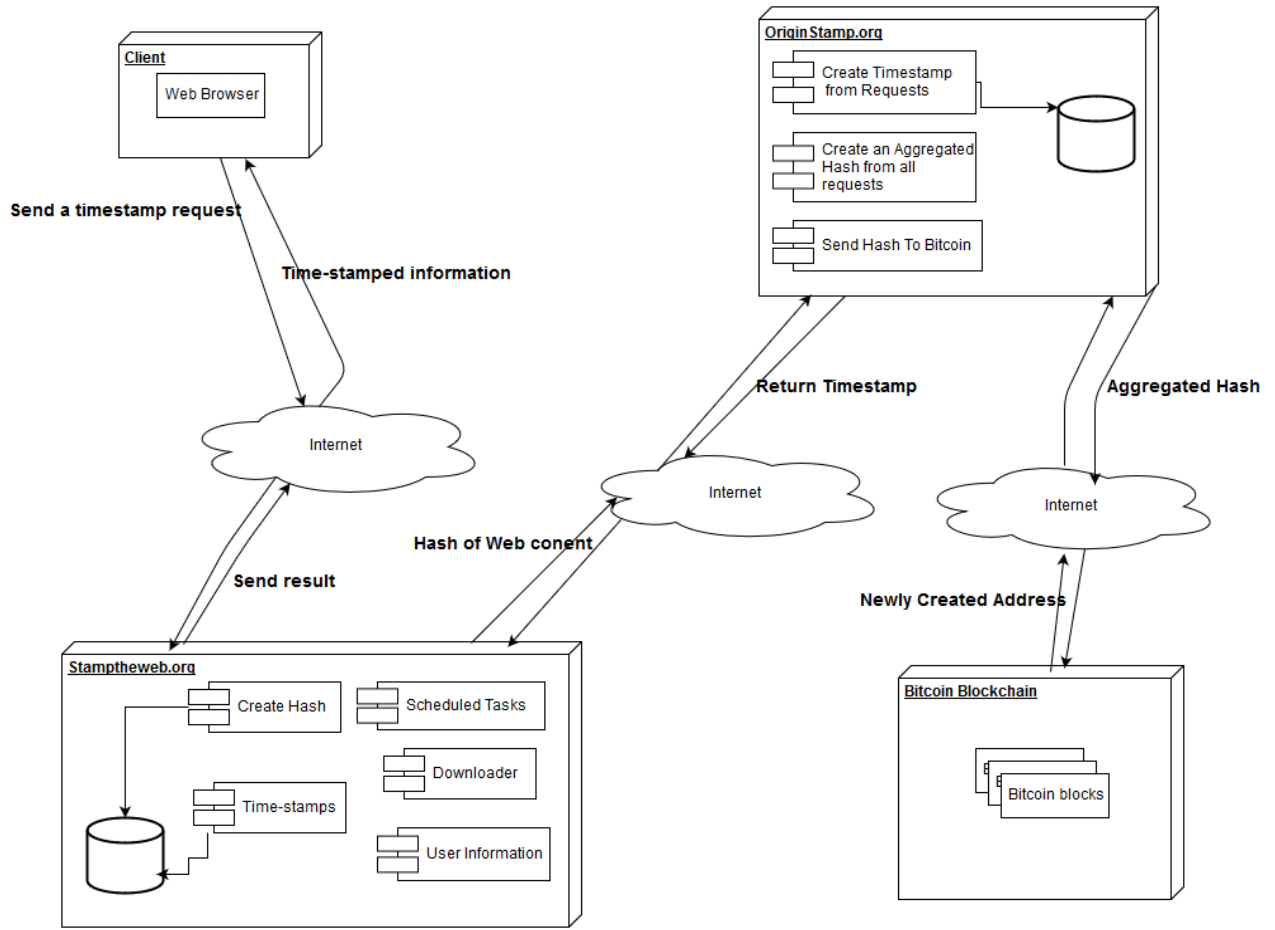

Figure 3: Deployment view of 'StampTheWeb.

### 3.1.1 Technologies Used

A range of technologies has been used in the development of front end of 'StampTheWeb. It is attempted to use most of the technologies online by providing the URL into the source. Some of the used technologies are as follows:

- JQuery
- JavaScript
- Bootstrap
- JSON
- D3
- Ajax

There are different files, which have also been used on the server locally, for example JavaScript and JQuery files. There are some D3 supported maps in the system, which require SVG images into the HTML files. The coordinates of these SVG images have been read using JSON files. These files also have been changed according to the user instructions into the python code at the server side. Some new JSON files have been created, depending on the specific requirements and instructions given by the user. Tab separated and comma separated files have been used for reading some data, i.e. proxies of different countries in order to open a URL from different locations.

Besides using general purpose technologies, some code for special purposes have been used, for example bootstrap calendar[17]. This calendar is used to show the calendar view of different time-stamped pages, either by the result of a search or through selected domains from the system. There is also an active community for this calendar, any help or bug fixing could be found from GitHub[18]. Moreover, 'simple map' in D3[19] modules have been used for creating maps for statistics, locations of proxies and showing block map of a provided URL.

---

# 4. System Design

If we divide both blueprints as discussed above in the system architecture into two different modules, then some of the used sub modules and their dependencies are explained in the module view of the system. Since the system is under development, this diagram may change over time. As the rule of a module view diagram, a dotted line represents the dependencies on other sub-modules, as shown in figure 4. In some cases, it would not be easy to find code for some modules. However, this division divides the sub-tasks into a diagram. To help understand the set of various functionalities within the system, each individual module is explained as follows:

**Login:** Contains all the activities related to logging in the user.
**Confirmation:** Confirmation activities such as confirmation token, and confirming after the email verified.
**Registration:** Registration activities such as sending email for completing registration.
**Management Functions:** Account management functions such as change in email, change in password, etc.

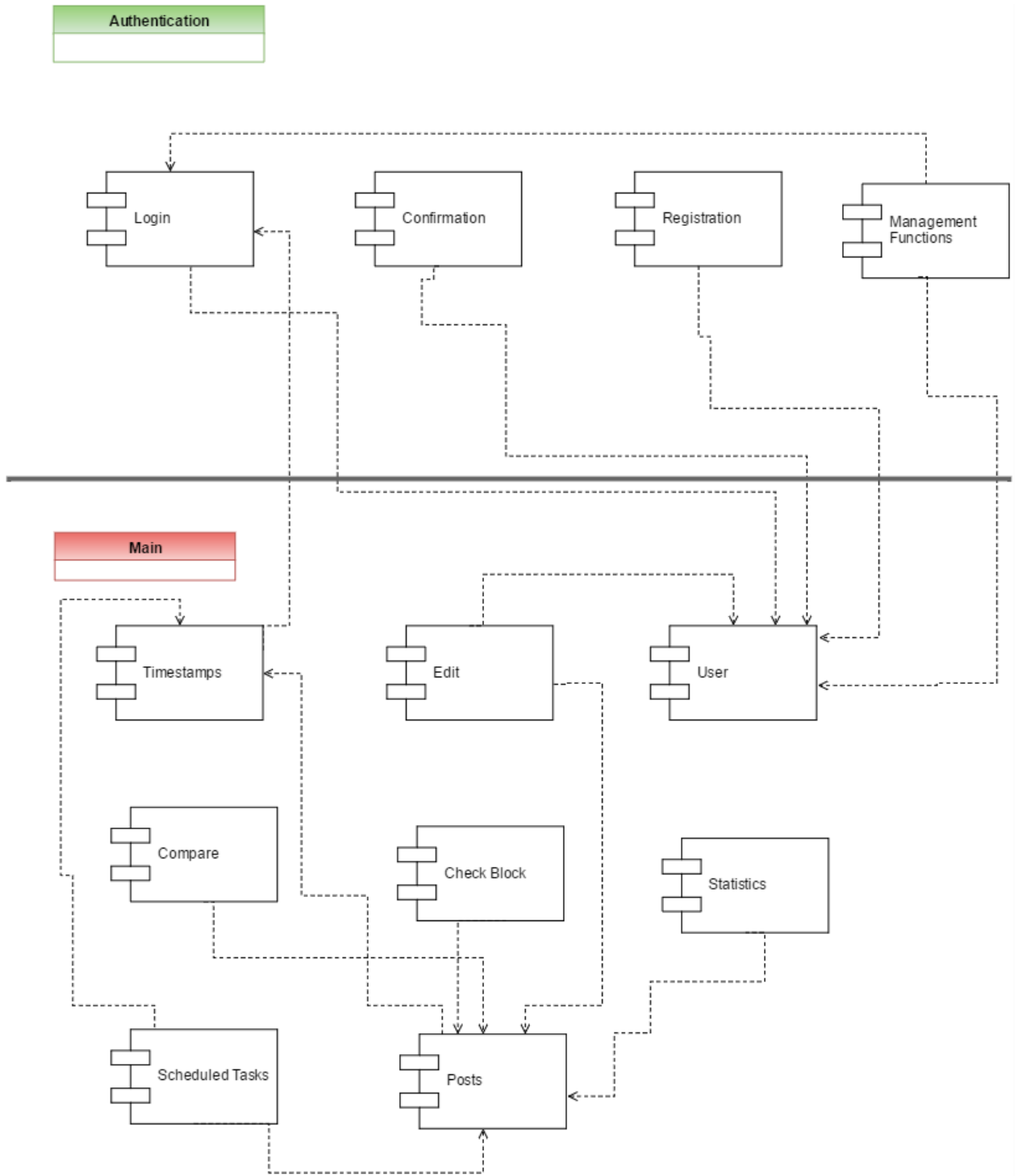

Figure 4: Modular View 'StampTheWeb.

**Time-stamps:** Creating time-stamps from the text in the URL with the help of 'OriginStamp'. It also contains creating pdf and screenshots.

**Edit:** Edit operations for the articles and user.

**Users:** All user related activities.

**Compare:** compare operations for different posts (with different countries or with each other).

**Check Block:** To check if post is blocked and all other blocked operations for selected articles.

**Statistics:** Activities related to create overall statistics visualization.

**Scheduled Tasks:** Activities, which need to be performed with a regular time interval, i.e. after every 86400 seconds.

**Posts:** Activities related to posts.

## 4.1 Database Design

In order to efficiently store data in a database, the database is divided into many tables and tables have been normalized to help prevent data redundancies. The database design of 'StampTheWeb is shown in figure 5. Each table is illustrated in detail as follows:

**location:** This table stores the locations of web URLs. First we get the public IPs of the URLs visited in the system, and then from public IP we obtain locations (by using free geoip[20] ) in order to visualize the overall statistics. Since it takes some time to take location from an IP, i.e. if the location is already present in the location table, the query is not sent to the free geoip API but rather obtained locally in order to save time.

**posts:** The table 'posts' contains all the information of a time-stamped page. If a web page is requested to be time-stamped again, the system checks the hash field in this table. If the hash is already present in the table, it means the URL is provided again and there is no change in the content of the provided URL. If hash is not present it means a new URL is provided or there are some changes in the content of the provided URL. System does not create multiple entries of an article, as it creates redundancies in the database.

**regular:** regular table corresponds to the scheduled tasks. This table is in 'one to many' relationship with the posts, table. Since one post can have multiple scheduled tasks, maybe by different users. Again at the time of a 'regular' time-stamp creation if a hash is present in the system, just another entry is created in the regular table. A record in the 'posts' table is not created.

**block:** This table contains records of articles, which are blocked in the selected countries. This table is with 'one-to-many' relationship with posts. One post can be blocked in many countries.

---

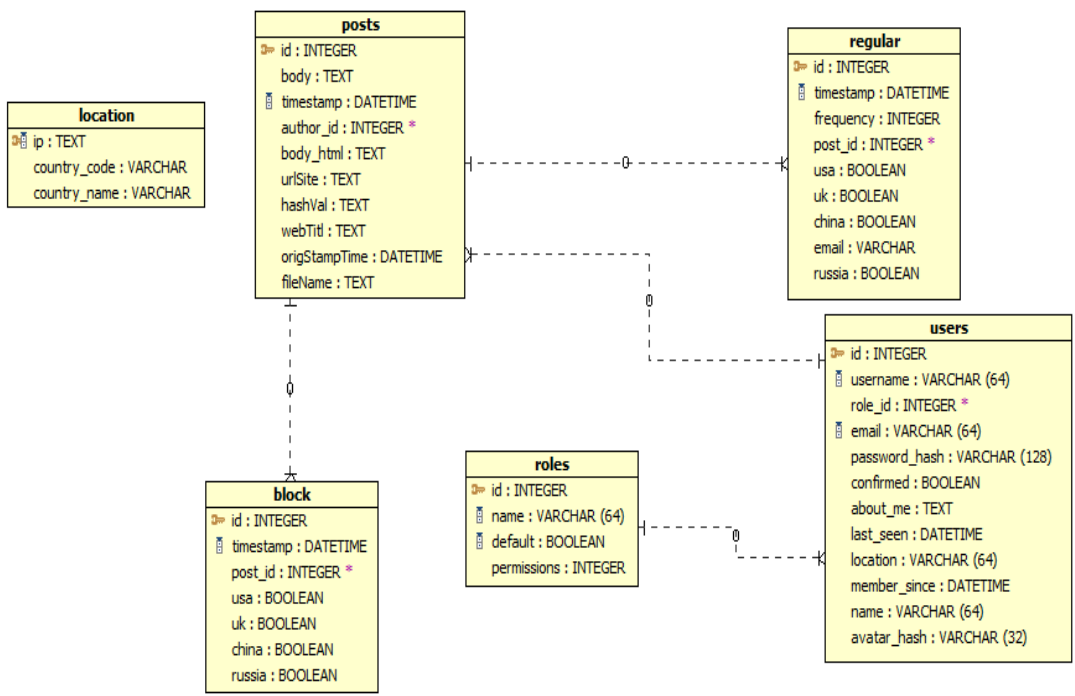

Figure 5: Entity relationship diagram 'StampTheWeb.

**roles:** roles table contains information regarding roles for a user. There are three roles added in the system Administrator, Moderator and User. As soon as a user creates an account a 'User' is automatically assigned to the new user. The other two types of account are for maintenance purposes. However, the roles can be extended. The roles table is also 'one-to-many' relationship with users table, since one user can have many roles. The roles are described in details in figure 6, which explains the use case diagram.

**users:** the 'users' table contains information of all the users in the system. This table is with 'one-to-many' relationship with posts. Since one user can have multiple posts. As discussed earlier a post is a time-stamped web article in the system.

## 4.2 Use Case Design

Every user registered in the system is a simple user. The other roles are just for configuration and administration. As shown in figure 6 use case diagram. A user, who is registered, can do everything available within the system. However, a user can only update her/his personal information. In order to change information of other users (such as profile, account) an administrator user is required. Similarly, a 'Moderator' can also do what a user can do. A moderator can also modify the content of other users such as posts. Moreover, in the future, if other features are introduced in the system, such as comments, then a moderator may also be able to change, update or delete it or the content of other users. A 'moderator' cannot update account information of a user, only an administrator can.

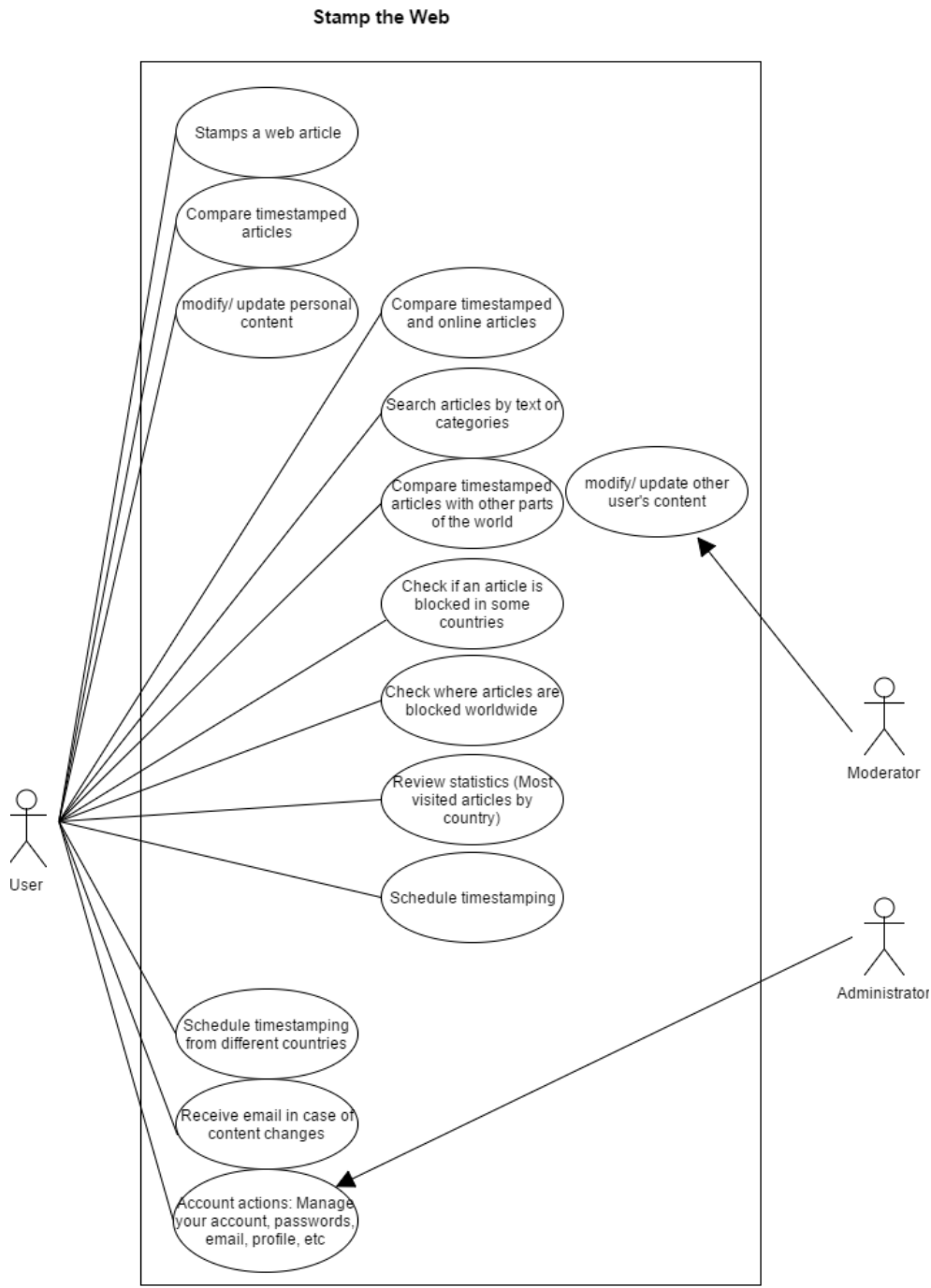

Figure 6: Use case diagram.

### 4.2.1 User Roles

The user roles are stored in the 'roles' table. The 'roles' table contains a permission column which defines each individual role of a user. This column contains a number, which would be read as binary string of eight digits. Each role is set according to this permission number. For example, an administrator has all 8 bits on, since it has access to everything in the system. Similarly, a moderator has the eighth binary bit on, which indicates access to content made by other users. Finally, a normal user, who is registered through registration in the result of 'sign up', has the fourth binary bit on. The fourth bit indicates writing access to the system. This is shown as follows:

Administrator: 11111111
Moderator:     00001000
User:          00000100

These roles are stored in the database as their decimal equivalent in 'StampTheWeb' as follows:
Administrator:        255
Moderator :    15
User:          7

For each user it is required to assign the 'role_id' the primary key of 'roles' table, in order to assign that user, the corresponding role. The 'user' role is by default assigned to the user, which has 'role_id' as '3'. User roles have been already defined in the code, in 'Roles' class present in 'models.py' file. In order to transfer user roles from code into the database for efficient usage, it is required to run the following command from the command prompt.

```
(venv) $ python manage.py shell
>>> Role.insert_roles()
```

In order to view all the roles present.

```
>>> Role.query.all()
[<Role u'Administrator'>, <Role u'User'>, <Role u'Moderator'>]
```

This way we can also define more roles by adding them into 'insert_roles()' method. This idea of setting up roles has been taken from the 'flask web development book'. [21]

## 4.3 Documentation Standards

It is attempted to arrange files and code in a reasonable manner. A vast range of files have been used, the most prominent are python files '.py' and html files '.html'. Files have been arranged according to the module they belong to. If a backend scripting file were a part of 'main_blueprint' it would be present in that folder. Moreover, html files are present in the 'templates' folder. There are further sub-sections created inside the 'templates' folder, in order to analytically divide the files in appropriate folders. For example, templates for sending emails to a user in case of change in content are in 'templates/mail' folder. Moreover, configuration and management scripts have been placed in the 'root' of the application. The code is self-explanatory and names of the files indicate the functionality of the files. Moreover, comments have been used with the classes and methods in order to help other developers understand the code.

---

[21] Grinberg, Miguel. *Flask Web Development: Developing Web Applications with Python.* " O'Reilly Media, Inc." 2014.

## 4.4 Naming conventions

Python naming conventions have been used i.e. PEP8[22]. It is encouraged to use variables methods and class names according to python naming conventions in the future or as follows:

```python
a_new_variable = 0
a_new_method():
class NewClass():
    def __init__(self):
```

## 4.5 Programming Standards

PEP8 codes style has been used while writing the code of python for server side scripting. The variables, methods and classes names follow pep8 naming conventions. Auto formatting has also been done using 'Pycharm' as well as PEP8 having been used in 'Pycharm' to automatically indicate if some code is not according to the coding standards of python.

## 4.6 Software Development Tools

It is encouraged to use 'Pycharm' or any other python integrated development environment (IDE). In order to better manage code and different set of files included in the project. The backend code is easily debugged using an IDE, however, some code is also present in the form of JavaScript and JQuery, which requires a modern web-browser, like Mozilla Firefox or Google Chrome to debug. Moreover, templating language Jinja2 also contains some scripting, however, it is understandable code and does not require any debugging. The application was initially developed using normal text editors, later 'Pycharm Community Edition' was used, which really helps manage the code, and makes development fast.

---

[22] "PEP 8 -- Style Guide for Python Code | Python.org." 2014. 30 Jul. 2016
<https://www.python.org/dev/peps/pep-0008/>

-

# 5. Component Description

Flask framework makes the development of the system very easy, but it is still required to have good understanding of the different set of used components. Some of the used modules are listed below:

## 5.1 Flask-SQLAlchemy

SQLAlchemy is the Python SQL toolkit and 'Object Relational Mapper' [23]. SQLAlchemy helps application developers write sophisticated code without writing SQL statements. The greater advantage of using it is not to worry about the backend databases. As soon as configuration parameters have been changed the SQLAlchemy starts using the databases whose configurations have been provided. Moreover, it is just required to change the model of the database from the code. SQLAlchemy will itself manage the model of database and communicate with database with the new model. The programmer does not require changing the database design from the database itself. Similarly, in our case it is required to change the 'model' file present at the app folder of the application, i.e. if it is required to change the 'users' table's model in the database. The following user class requires change in the code. Everything else would be managed by SQLAlchemy.

```python
class User(UserMixin, db.Model):
    __tablename__ = 'users'
    id = db.Column(db.Integer, primary_key=True)
    email = db.Column(db.String(64), unique=True, index=True)
    username = db.Column(db.String(64), unique=True, index=True)
    role_id = db.Column(db.Integer, db.ForeignKey('roles.id'))
    password_hash = db.Column(db.String(128))
    confirmed = db.Column(db.Boolean, default=False)
    name = db.Column(db.String(64))
    location = db.Column(db.String(64))
    about_me = db.Column(db.Text())
    member_since = db.Column(db.DateTime(), default=datetime.utcnow)
    last_seen = db.Column(db.DateTime(), default=datetime.utcnow)
    avatar_hash = db.Column(db.String(32))
    posts = db.relationship('Post', backref='author', lazy='dynamic')
    very = db.relationship('Verify', backref='author', lazy='dynamic')
```

Integrating SQL scripts into the code is very easy. It is required to use the built-in SQLAlchemy methods in order to perform SQL operations into the code. There is no need to write raw SQL statements. Given below are some examples, which may be later used:

---

[23] "SQLAlchemy - The Database Toolkit for Python." 2005. 31 Jul. 2016
<http://www.sqlalchemy.org/>

Adding a new record in a model class.

```
new_record = Model_Class(value_1=<value_1>, value_2=<value_2>
............,value_n=<value_n>, foreign_key=<foreign_key_object>)
db.session.add(new_record )
db.session.commit()
```

Querying from the database
```
result = Model_Class.query.get_or_404(search_value)
```

Performing SQL operations (AND)
```
result=
Model_Class.query.filter(and_(Model_Class.column_1.like(value_1)
,
Model_Class.column_2.like(value_2))).first()
```

## 5.2 Flask-Migrate

Flask migrate helps change the database structure without affecting the existing data present in the database. After the change in the database model, it is required to let Flask-Migrate upgrade the database and effect the changes in the 'mode' into the database. Migration can be done using 'manage.py' file present at the root of the application. The python commands for database migration are as follows:

In order to create a migration folder:
```
(venv) $ python manage.py db init
```

In order to create migration scripts and version in the migration folder:
```
(venv) $ python manage.py db migrate
```

In order to apply changes into the database without affecting the data:
```
(venv) $ python manage.py db upgrade
```

All the migration scripts have already been created. In the case of change in the database 'model', only second and third commands needs to be executed.

## 5.3 Flask Bootstrap

Bootstrap is an HTML, CSS and JavaScript framework for developing a responsive web project. If a web project contains Bootstrap integration, it can automatically adjust itself on the user requested screen, mobile tablet computer, etc. It is not required to develop different front ends based on the user request, if bootstrap is used. Different messages used in 'StampTheWeb also use Bootstrap messaging. Bootstrap also helps us write less JavaScript code by just calling different Bootstrap classes. In order to add a new front end module, it is necessary to extend the 'bootstrap/wtf.html' web page in the following way:

```
{% import "bootstrap/wtf.html" as wtf %}
```

This way we can use all the Bootstrap functionalities and components. [24]

## 5. Future System Improvements

It is attempted to make code reusable as much as possible, by using methods and classes. However, Flask applications have a different approach. The code present inside the 'blueprints' can only be used as soon as the 'application context' is available. Some code used to send email is present in the 'main' module of the application. This code cannot be re-used for sending emails for other purposes, like automatic email sending to user, if there is a change in the time-stamped content. In addition, if a requested page is blocked in a country of user's choice, the system sends an email to the associated user. This code executes once after 84600 seconds, and it is found that flask application context is not present and we cannot reuse this code from the 'main_blueprints'. Hence the code is re-written again in 'send_email' file. Other than this issue no other code is re-written, and it should be attempted to use as little code as possible.

In case of getting text from web articles we are using a simple python 'requests' module. This module works fine for different web articles, Wikipedia and other news sources. We get the text from a user's provided URL as follows:

First we get the response from the user provided URL
```
res = requests.get(url)
```

Then convert the text present at the URL into a document.

```
doc = Document(res.text)
```
Then we simply get a document's summary by using the appropriate method, and then we convert this summary as an HTML document on the server, for later comparison with web pages.

```
text = '<!DOCTYPE html PUBLIC "-//W3C//DTD XHTML 1.0 Transitional//EN"
"http://www.w3.org/TR/xhtml1/DTD/xhtml1' \
```

---

```
    '-transitional.dtd">\n' + '<head>\n' + \
    '<meta http-equiv="Content-Type" content="text/html" ' \
    'charset="' + encoding + '">\n' + '</head>\n' + '<body>\n' \
    + '<h1>' + doc.title().split(sep='|')[0] + '</h1>'

text += doc.summary() + '</body>'
```

This approach works, but in some cases some text from web pages is missing. It is required to get complete text from the given URL. One possible approach could be as follows:

Rather than converting the text from a URL into a doc, we can simply get the text from the response of a URL as:

```
text = res.text
```

This text also contains all the html tags and links, which needs to be removed when text is compared later with other versions. More importantly, it is required to save HTML using appropriate encoding utf-8 in case of the server being a windows system.

There are also other approaches to get text from HTML. One approach is using the module 'beautiful soup 4' (bs4). Initially, bs4 was used, however, extracting text using bs4 took more time. Therefore, the use of bs4 was removed from the code.

# 6 Installation and Outlook

'StampTheWeb requires python 3 and Flask to run. It was tested with the following flask and python packages. In order to run application smoothly, following packages are necessary :

- alembic==0.8.6
- Babel==1.3
- beautifulsoup4==4.4.1
- bleach==1.4.3
- blinker==1.4
- chardet==2.3.0
- click==6.6
- cryptography==1.2.3
- cssselect==0.9.1
- docutils==0.12
- dominate==2.2.0
- enum34==1.1.2
- feedparser==5.1.3
- Flask==0.11.1
- Flask-Bootstrap==3.3.6.0
- Flask-Login==0.3.2
- Flask-Mail==0.9.1
- Flask-Migrate==1.8.0
- Flask-Moment==0.5.1
- Flask-PageDown==0.2.1
- Flask-Script==2.0.5
- Flask-SQLAlchemy==2.1
- Flask-SSLify==0.1.5
- Flask-WTF==0.12
- funcsigs==0.4
- gdata==2.0.18
- gevent==1.1.1
- greenlet==0.4.9
- gunicorn==19.4.5
- html5lib==0.9999999
- idna==2.0
- ipaddress==1.0.16
- itsdangerous==0.24
- Jinja2==2.8
- linecache2==1.0.0
- lxml==3.6.0
- Mako==1.0.4

- Markdown==2.6.6
- MarkupSafe==0.23
- mock==1.3.0
- pbr==1.8.0
- Pillow==3.1.2
- psutil==3.4.2
- psycopg2==2.6.1
- pyasn1==0.1.9
- PyChart==1.39
- pycrypto==2.6.1
- pydot==1.0.29
- Pygments==2.1
- pyinotify==0.9.6
- PyOpenGL==3.0.2
- pyOpenSSL==0.15.1
- pyparsing==2.0.3
- Pyrex==0.9.8.5
- python-dateutil==2.4.2
- python-editor==1.0
- python-ldap==2.4.22
- python-openid==2.2.5
- python-stdnum==1.2
- pytz==2014.10
- PyWebDAV==0.9.8
- PyYAML==3.11
- readability-lxml==0.6.2
- reportlab==3.3.0
- requests==2.10.0
- roman==2.0.0
- schedule==0.3.2
- simplejson==3.8.1
- six==1.10.0
- SQLAlchemy==1.0.13
- traceback2==1.4.0
- unittest2==1.1.0
- unity-lens-photos==1.0
- uTidylib==0.2
- vatnumber==1.2
- virtualenv==15.0.2
- visitor==0.1.3
- vobject==0.8.1rc0
- Werkzeug==0.11.10
- WTForms==2.1

- xlwt==0.7.5
- ZSI==2.1a1

To install a python package use the following command:
```
$ pip install [python-package]
```

A 'requirements.txt' file can also be used to install these packages, which is present at the root of the source code. In order to install all the packages from 'requirements.txt', pip command should be followed.

```
$ pip install -r requirements.txt
```

To clone this source locally into your computer:
```
$ git clone https://github.com/Sebisnow/StampTheWeb.git
```

To run the source code locally:
```
$ python manage.py runserver
```

## 6.1 Limits for DB

Currently, initial DB is created using SQLite. SQLite preserves two Terabytes of data. This data storage excludes PDF and screenshot data storage. It is just user and their post information storage, which would be saved in a DB. We could also switch to another DB that just needs change in the configurations. Only environment variables for DB configurations need to be changed. If configurations for another database are given, the system switches itself to that database automatically.

Currently, no database username and password has been used, however, variables for username and password could be added as new environment variables in the code, if the other database contains username and password.

## 6.2 Good Practice for Environment Variables

For better configurations it is suggested to use environment variables rather than writing them into the code. If environment variables are used and values for variables used in the code are not hardwired in the code, it saves time in configurations. The setting of environment variables, as shown below, help us to change any value using a command line and does not require making changes in the code and updating the code commit and then commit changes on the server.

```
SECRET_KEY = os.environ.get('SECRET_KEY') or 'hard to guess string'
SQLALCHEMY_COMMIT_ON_TEARDOWN =
os.environ.get('SQLALCHEMY_COMMIT_ON_TEARDOWN ') or True
MAIL_SERVER = os.environ.get('MAIL_SERVER') or 'smtp.gmail.com'
```

```
MAIL_PORT = os.environ.get('MAIL_PORT') or 587
MAIL_USE_TLS = os.environ.get('MAIL_USE_TLS') or True
MAIL_USERNAME = os.environ.get('MAIL_USERNAME') or '<username>'
MAIL_PASSWORD = os.environ.get('MAIL_PASSWORD') or '<password>'
STW_MAIL_SUBJECT_PREFIX = os.environ.get('STW_MAIL_SUBJECT_PREFIX') or
'[StampTheWeb]'
STW_MAIL_SENDER = os.environ.get('STW_MAIL_SENDER') or
'<stamptheweb@gmail.com>'
STW_ADMIN = os.environ.get('STW_ADMIN') or 'stamptheweb@gmail.com'
STW_POSTS_PER_PAGE = os.environ.get('STW_POSTS_PER_PAGE') or 20
STW_CHINA_PROXY = os.environ.get('STW_CHINA_PROXY') or
"101.201.42.44:3128"
STW_USA_PROXY = os.environ.get('STW_USA_PROXY') or "169.50.87.252:80"
STW_UK_PROXY = os.environ.get('STW_UK_PROXY') or "89.34.97.132:8080"
STW_RUSSIA_PROXY = os.environ.get('STW_RUSSIA_PROXY') or
"80.240.114.77:8000"
CHINA_PROXY = os.environ.get('CHINA_PROXY') or "60.216.40.135"
USA_PROXY = os.environ.get('USA_PROXY') or "199.115.117.212"
UK_PROXY = os.environ.get('UK_PROXY') or "90.216.222.23"
RUSSIA_PROXY = os.environ.get('RUSSIA_PROXY') or "80.240.114.77"
SERVER_URL = os.environ.get('SERVER_URL') or 'https://stamptheweb.org'
SQLALCHEMY_TRACK_MODIFICATIONS =
os.environ.get('SQLALCHEMY_TRACK_MODIFICATIONS') or False
```

## 6.2.1 Description for Environment Variables

| Name | Description |
|---|---|
| SECRET_KEY | The key for generating login session. By default, user sessions are stored in client-side cookies that are cryptographically signed using the configured SECRET_KEY. Any tampering with the cookie content would render the signature invalid, thus invalidating the session.[25] |
| SQLALCHEMY_COMMIT_ON_TEARDOWN | Set true if automatic commits of DB are required using SQLAlchemy |
| MAIL_SERVER | The mail server email address that requires sending emails to the user for different purposes. |

---

[25] Grinberg, Miguel. *Flask Web Development: Developing Web Applications with Python.* " O'Reilly Media, Inc." 2014.

| | |
|---|---|
| **MAIL_PORT** | The port for SSL or TLS that is being used by the mail server. |
| **MAIL_USE_TLS** | 'True' if Mail server uses TLS protocol. |
| **MAIL_USERNAME** | Username of email account used to send emails to the users. |
| **MAIL_PASSWORD** | Password of email account used to send emails to the users. |
| **STW_MAIL_SUBJECT_PREFIX** | The subject prefix of each mail sent by the system. |
| **STW_MAIL_SENDER** | Mail sender's 'email address' of the system. |
| **STW_ADMIN** | Admin email address of the system. |
| **STW_POSTS_PER_PAGE** | How many posts need to be displayed on a single page? |
| **STW_CHINA_PROXY** | A valid public IP address of a proxy of China. To load a webpage from China. |
| **STW_USA_PROXY** | A valid public IP address of a proxy of USA. To load a webpage from USA. |
| **STW_UK_PROXY** | A valid public IP address of a proxy of UK. To load a webpage from UK. |
| **STW_RUSSIA_PROXY** | A valid public IP address of a proxy of Russia. To load a webpage from Russia. |
| **CHINA_PROXY** | Public IP of the China proxy without the port number. This is to get the location of the proxy. |
| **USA_PROXY** | Public IP of the USA proxy without the port number. This is to get the location of the proxy. |
| **UK_PROXY** | Public IP of the UK proxy without the port number. This is to get the location of the proxy. |
| **RUSSIA_PROXY** | Public IP of the Russia proxy without the port number. This is to get the location of the proxy. |
| **SERVER_URL** | The URL of the server currently used. In order to send scheduled emails to the user, e.g. 'https://stamptheweb.org' |

| **SQLALCHEMY_TRACK_MODIFICATIONS** | Set 'True' if system should generate URL automatically for sub domains. However, in our case, some problems found, so it is set 'False' |
| --- | --- |

## 6.3 Reporting Bug section for source code

For reporting bugs and issues in the system a 'GitHub' issue tracker has been set. It is very important for working in teams and collaboratively solving issues. 'GitHub' provides a sophisticated issue tracking system associated with the source code. This issue tracking can be accessed using:
Stamp The Web Issue tracking: URL:
https://github.com/Sebisnow/StampTheWeb/issues

## 6.4 Future Improvements

### 6.4.1 Articles by modifications

The system may include statistics for articles that have a high modification rate once time-stamped in the system. Then the user could discover which domains/ media outlets tend to modify their reporting, maybe also their political opinion.

### 6.4.2 Help

Help and reporting bugs sections are not included in the system. It is necessary that the user is able to find some components in the system. A help section may also contain videos that visualize certain tasks in the system, this can also make system very helpful.

### 6.4.3 Consistent Hash

It is noticed with some web articles that their hash is not consistent, even if the content of the web article is not changed. When a web article is compared with another version, even if there is no change in the web article, the system finds a changed hash, hence a flashing message appears 'Change in content found'. This needs to be fixed.

### 6.4.4 Search for Blocked Articles

Currently, a search works for all the time-stamped articles. We can simply browse blocked web articles by going to 'where it is blocked' page. We also need to add search option for blocked articles.

### 6.4.5 Check Multiple proxies

In order to find if a web page is blocked in a country, we only check one proxy. If that proxy is unable to find the requested page, it is deemed as blocked in that country. However, sometimes a proxy doesn't work and in that case deeming the page blocked is false. We need to add more than one proxy, at least three for a country. In order to add more proxies, the current code needs to be replaced with the following:

```
proxy1 = '169.50.87.252:80'
proxy2 = '89.34.97.132:8080'
proxy3 = '11.22.33.44:80'
requests.get(url, proxies={"http":proxy1,"http":proxy2
,"http":proxy3})
```

### 6.4.6 Search Engine Optimization

The foremost way for marketing a system is to bring it into the front of search engine results. Yet the use of keywords in the front end of the system should be done in such a way that it remains search friendly to the search engine. For example, useful keywords in the header as well as in the body section of an HTML page. However, the system has been shown in search engine listings as shown in **Error! Reference source not found.**. Whereas, it is required to make search engine optimization with different keywords, in order to outreach maximum number of users. Hence, it is required to optimize system for search engines.

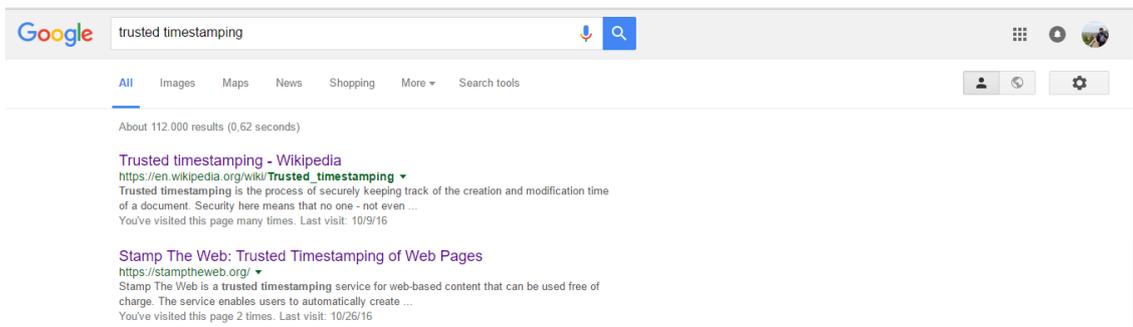

Figure 7: 'StampTheWeb' shown on second position when the word 'trusted time-stamping' has been searched

# Document Control

**Title:**          Stamp the Web Developer's Documentation

**Issue:**          Issue 1

**Date:**          10 August 2016

**Author:**          Waqar Detho

**Filename:**          technical_doc

## Document Signoff

| Nature of Signoff | Person | Signature | Date | Role |
|---|---|---|---|---|
| Authors | Waqar Detho | | | Software Developer |
| Reviewers | | | | |

## Document Change Record

| Date | Version | Author | Change Details |
|---|---|---|---|
| 10 August 2016 | Issue 1 Draft 2 | Waqar Detho | First complete draft |
| 4 December 2016 | Issue 2 Draft 2 | Waqar Detho | Review for Thesis |
| | | | |
| | | | |

## 8.4 Data and Source-code Downloads

**Source code 'StampTheWeb' with installation instructions:**

https://github.com/waqaralidetho/finalSourceCode

**Technical Documentation 'StampTheWeb':**

https://github.com/waqaralidetho/documentation

**Screen Recordings of Usability Testing Sessions:**

https://www.youtube.com/watch?v=YU8eUayGS50

https://www.youtube.com/watch?v=7T6x2pWJf-Q

https://www.youtube.com/watch?v=avk3dU3Au6E

https://www.youtube.com/watch?v=0ORDvmZ9BA4

https://www.youtube.com/watch?v=CQqTwD0ANC4

https://www.youtube.com/watch?v=7iHLSfiINww&feature=youtu.be

https://www.youtube.com/watch?v=nz0ntNpGqlI

https://www.youtube.com/watch?v=_sgOxhY_cSM

**YouTube Channel contains all above videos (Waqar Detho)**

https://www.youtube.com/channel/UCzOu54h-KPjvbSi82iidkAg